\documentclass[sigconf]{acmart} 

\usepackage[utf8]{inputenc}
\usepackage{multirow}
\usepackage{booktabs}
\usepackage{graphicx}
\usepackage{subcaption}
\usepackage{xcolor}
\usepackage{enumerate}

\setcopyright{none}
\renewcommand\footnotetextcopyrightpermission[1]{} 
\AtBeginDocument{%
  \providecommand\BibTeX{{%
    \normalfont B\kern-0.5em{\scshape i\kern-0.25em b}\kern-0.8em\TeX}}}
    
\acmConference[UIST '22]{The 35th Annual ACM Symposium on User Interface Software and Technology}{October 29-November 2, 2022}{Bend, OR, USA}



\newcommand{\rebut}[1]{{\color{black}#1}}

\title[Personalized Game Difficulty Prediction Using Factorization Machines]{Personalized Game Difficulty Prediction\\Using Factorization Machines}

\author{Jeppe Theiss Kristensen}
\email{jetk@itu.dk}
\orcid{0000-0002-3145-4991}
\affiliation{%
  \institution{IT University of Copenhagen}
  \department{Digital Design}
  \city{Copenhagen}
  \country{Denmark}
}

\author{Christian Guckelsberger}
\email{christian.guckelsberger@aalto.fi}
\orcid{0000-0003-1977-1887} 
\affiliation{%
  \institution{Aalto University}
  \department{Department of Computer Science}
  \city{Espoo}
  \country{Finland}
}

\author{Paolo Burelli}
\email{pabu@itu.dk}
\orcid{0000-0003-2804-9028}
\affiliation{%
  \institution{IT University of Copenhagen}  \department{Digital Design}
  \city{Copenhagen}
  \country{Denmark}
}

\author{Perttu Hämäläinen}
\email{perttu.hamalainen@aalto.fi}
\orcid{0000-0001-7764-3459}
\affiliation{%
  \institution{Aalto University}
  \city{Espoo}
  \country{Finland}
}

\date{November 2022}

\begin{document}

\begin{abstract}
    The accurate and personalized estimation of task difficulty provides many opportunities for optimizing user experience. However, user diversity makes such difficulty estimation hard, in that empirical measurements from some user sample do not necessarily generalize to others.

    In this paper, we contribute a new approach for personalized difficulty estimation of game levels, borrowing methods from content recommendation. Using factorization machines (FM) on a large dataset from a commercial puzzle game, we are able to predict difficulty as the number of attempts a player requires to pass future game levels, based on observed attempt counts from earlier levels and levels played by others. In addition to performance and scalability, FMs offer the benefit that the learned latent variable model can be used to study the characteristics of both players and game levels that contribute to difficulty. We compare the approach to a simple non-personalized baseline and a personalized prediction using Random Forests. Our results suggest that FMs are a promising tool enabling game designers to both optimize player experience and learn more about their players and the game.
    
\end{abstract}

\maketitle


\pagestyle{fancy} 

\section{Introduction}
\label{sec:intro}
Understanding and estimating task difficulty is a fundamental problem in Human-Computer Interaction (HCI). While good user interface design aims to minimize the difficulty of completing tasks and achieving goals, we might also be interested in introducing challenges of just the right difficulty -- neither too easy nor too hard -- to e.g.~support learning \cite{wilson2019eighty} or create enjoyable video games \cite{sweetser2005gameflow, chen_flow_2007}. 
Estimating how difficult a challenge is for the target user is hard due to the diversity of factors affecting difficulty. On a macro level, the difficulty of making decisions can be boiled down to skill, the available time, and the inherent difficulty of the decision \cite{anderson2017assessing}. However, only the available time is straightforward to measure in general, and in HCI, executing decisions can pose additional perceptual-motor difficulties.



This paper investigates difficulty in the particular setting of casual mobile puzzle games where the player progresses in the game by completing discrete challenges, or levels. In this context, a common operationalization of difficulty is the chance of the player completing the level, or pass rate. This is typically calculated as the count of successful attempts divided by the total attempts on the level. 
The inverse represents the average number of attempts per complete. 
Averaged over all players, this is a useful difficulty metric for game designers since it gives a tangible measure of how much time a player spends on a level before proceeding, which can help identify where the player might feel stuck \cite{drey_be_2021} as a potential cause for churn. However, a significant drawback of this measure is that it does not account for individual differences in skill. Simply looking at the aggregate population, therefore, runs the risk of alienating anybody but the average player who, in many cases, is neither representative nor the most important prediction target. A more useful approach for game designers should thus take both individual player and level information into account.


This challenge of accounting for individual difficulties has been approached in multiple ways. 
A common practice is to utilize \emph{dynamic difficulty adjustment} (DDA) in which the predictions are used to automatically adjust game parameters (e.g. number of enemies) and tailor the game difficulty to maximize aspects such as retention or monetization of individual players.
However, this kind of fine-grained control over the levels is not always possible due to technical aspects (difficult to implement, uncertainty about parameters' effect on difficulty, etc.) or level and game design choices (levels requiring a specific strategy or visuals, difficulty curve must follow a certain pattern, etc.). Moreover, fully automated difficulty adjustment might not be attractive for game designers if they want to retain some control over the player experience. An alternative approach that can be of practical use should therefore also capture and explain player-level interactions and generate knowledge for the game designers that allows them to be more proactive.

The main objective of this paper is to showcase such a framework that game designers can use to both understand the interaction between players and levels and to estimate level difficulty for individual players. 
\rebut{Rather than updating the estimates between game rounds, as is common in many DDA approaches, the goal of this framework is to leverage daily play session data to inform the offline work of level designers and help them understand the player base.} 
To achieve this, we use \textit{factorization machines} (FMs) which are especially known from recommendation systems \cite{rendle2010factorization,rendle2011fast,rendle_factorization_2012,hong2019interaction}. FMs allow predicting user-content interactions by estimating latent variables that describe each user and each piece of content.


To better understand how FMs can be applied for difficulty estimation, we investigate the following research questions:

\begin{enumerate}
    \item[\textbf{RQ1:}] How do FMs compare to other difficulty prediction methods?
    \item[\textbf{RQ2:}] How many observations of a player are necessary before it is possible to discern them from the average player?
    \item[\textbf{RQ3:}] What do the FM model latent variables mean or represent? 
\end{enumerate}


\paragraph{Contribution} In summary, we examine FMs as a novel approach for personalized game level difficulty prediction. Using a large dataset of 700k players from the commercial puzzle game Lily's Garden, we compare FMs against both a naive non-personalized baseline and personalized predictions using Random Forests (RF). Our results support the use of FMs as a promising tool that clearly outperforms the other methods, especially if augmented with similar additional features as in the RF regression.

\section{Related Work}

To contextualize our contribution, we first resolve ambiguity around the concept of difficulty, and survey related work on operationalizing and quantifying difficulty in videogames. We then survey existing player difficulty prediction models.

\subsection{Game Difficulty}
\label{sec:operationalisation}
Game difficulty is a highly ambiguous concept \cite{denisova_challenge_2017}, with at least two meanings \cite{denisova_measuring_2020}. Firstly, it can denote an \emph{intrinsic attribute} of a game, characterizing a game-internal task based on its objective and the barriers that prevent potential players from achieving it. Secondly, it can describe a \emph{relational attribute} between the game and the player, characterizing the player's experience of the task based on their individual skill and history. Often, researchers use the notion of \emph{perceived difficulty} to convey the second, experiential meaning. To complicate things further, difficulty is often used synonymously with \emph{challenge}. However, we can draw a subtle distinction based on the concepts' \emph{valence} \cite{denisova_measuring_2020}: players typically consider a game task difficult, if it causes them frustration and discomfort; challenging tasks in contrast are stimulating and convey a feeling of being in control over the outcome \cite{lazzaro2004we}. Here, we primarily use the notion of difficulty, but without appealing to its negative valence.

Difficulty can be actively sought by players as a goal experience \cite{cairns2016engagement}, or it can form the foundation \cite{power2019lost} of other experiences. Famously, perceived difficulty contributes to player \emph{enjoyment}, as investigated by Alexander et al. \cite{alexander_investigation_2013}. Another such goal experience is \emph{flow} -- a state in which a player feels engulfed in the task and loses track of time and worries \cite{csikszentmihalyi1990flow, chen_flow_2007}. It results from exposing a player to difficulties that are optimal with respect to their individual skills. 
Self-determination theory \cite{Ryan2006,tyack2020self} posits that optimal difficulty satisfies players' intrinsic need for feelings of competence and in effect yields motivating gameplay. Amongst others, flow and intrinsic motivation contribute to player engagement \cite{brockmyer2009development}, which, similar to enjoyment, constitutes a core game design objective.

Given the many ways through which difficulty impacts player experience, it is unsurprising that game designers and researchers have sought ways to operationalize difficulty for use in design-time quality control or runtime optimization. In the particular case of puzzle games, Pusey et al. \cite{pusey_puzzle_2021} proposed several objective measures to quantify difficulty, including the number of incorrect attempts, the number of actions used, and the time taken to solve a puzzle. The average number of attempts that players spend to complete a level, or inversely the pass rate, has been used to assess difficulty in puzzle games such as Angry Birds \cite{roohi_predicting_2021} and Lily's Garden \cite{jeppe2020estimatinglearning}. A player's experienced number of successes and failures also constitutes a major component of perceived difficulty, as empirically identified by Denisova et al. \cite{denisova_measuring_2020} in the development of a questionnaire to assess perceived difficulty in games. Similar to previous work, we operationalize an individual player's perceived difficulty as the average attempts per complete.
Given that players cannot repeat levels in this study, this is equivalent to the number of attempts they require to complete a specific level.

\subsection{Predicting Difficulty}
\label{sec:related_work_models}
Previous research has approached player difficulty prediction in numerous ways depending on the application and research purpose.
Common application areas of difficulty prediction include DDA and automated playtesting for quality assurance. We next identify several shortcomings of existing work based on selected examples and highlight how our contribution overcomes them.

Existing approaches typically predict difficulty aggregated over a player population 
rather than individual experiences (e.g.~\cite{gudmundsson2018human,roohi_predicting_2021, jeppe2020estimatinglearning, mourato2014difficulty, kamaldinov2019deep,kreveld2015automated}). An example of this is the work by Gudmundsson et al. \cite{gudmundsson2018human} where an AI game-playing agent was used to extract a preliminary estimate of the level pass rate. This estimate was then combined with level features to create a binomial regression model to predict the overall pass rate on the level. We consider the focus on aggregated predictions a shortcoming, as the perceived difficulty is affected by individual skill and experience, and an aggregate prediction is thus likely to predict the various goal experiences that perceived difficulty contributes to less accurately. In contrast to existing approaches, this paper focuses on predicting individual player difficulty, thus moving one step closer to predicting perceived difficulty as a relational attribute between one player and the game. 

Existing studies that focus more on individual player difficulty prediction often suffer from practical and technical barriers: they are either operating on a very small scale (e.g. \cite{silva_dynamic_2015,sarkar2017level,wheat2016modeling,lora2016dynamic}), only consider toy problems (e.g. \cite{gonzalezduque2020findingerror}), or both (e.g. \cite{gonzalezduque2021fastmodelling,jennings2010polymorph}).
Gonzalez-Duque et al., \cite{gonzalezduque2021fastmodelling} for instance, use a Bayesian optimization approach to reliably estimate the time it would take a player to complete a Sudoku level given the number of pre-filled digits or a simple puzzle level given two level descriptors after having observed the player for 5 or 15 game rounds, respectively.
However, the limited number of play traces (<300) in each game lead to a large variance in the results. It is uncertain how the approach would scale in a live game with several player cohorts and continuous updates to the game. In our study, we use data from more than 700,000 players over a 6 month period to demonstrate the applicability of our approach to a large-scale, complex commercial game system.

A common approach in many DDA methods is to consider the recent history of players in order to adapt the game content. However, not all methods are able to take advantage of the wealth of information that is available about the players and levels (e.g. \cite{xue_dynamic_2017,li_difficulty-aware_2021, gonzalezduque2021fastmodelling,zook_skill-based_2012, bunian_modeling_nodate}).
For instance, Xue et al. \cite{xue_dynamic_2017} use a probabilistic graph that estimates the probability of winning, failing and churning based only on the player's progress and the current number of attempts; however, they do not leverage more descriptive features, such as average playtime, that have been found to be correlated with player engagement \cite{drachen_rapid_nodate,kristensen_combining_2019}.
In our approach, we aim to leverage a maximum range of player and level data, combined with high-cardinality data such as individual player-level interactions.

\rebut{Especially in the domain of educational games, it is common to explicitly model a learning curve and introduce new elements through tutorials to build up player competencies for solving more complex tasks later \cite{linehan_learning_2014}.
The Additive Factors Model~\cite{cen2006learning} is a popular method in this context for determining a player/student's chance of success on a task that requires certain skills~\cite{hardstead2015usinggame}.
However, many related approaches require labelling the required skills beforehand, and more specialized domain models can be hard to validate and generalize~\cite{goldin2018most}.
In this work, we do not assume any specific structure of player skill and knowledge and instead learn latent representations which we can interpret afterward.}

As last two shortcomings, we note that, especially in live game operations, it is not desirable to have a black-box prediction method that requires expert knowledge to operate \cite{mahlmann_predicting_2010}, as in related work that utilizes deep learning methods (e.g. \cite{pfau2020enemy,moon2020dynamic,or2021dl}). In addition, methods that require complex implementations (e.g. \cite{weber2020dynamic}) are also undesirable, as any unexpected behavior may be hard to troubleshoot and make game designers lose trust in the system.
Moreover, little to no knowledge is generated that enables game designers to make informed decisions about their games.
In this work, we demonstrate that factorization machines (FMs) afford ease of use, do not require any special input other than data about the player and the number of attempts they spent on a given level, and afford straightforward interpretation that game designers may be able to use for other tasks such as personalized offers or churn prediction.

To determine feasible approaches for difficulty prediction in games, it is worth looking beyond games.
A similar problem to difficulty prediction is student grade prediction \cite{sweeney_next-term_2016,ren_grade_2019}. Sweeney et al. \cite{sweeney_next-term_2016} use a number of methods to predict the grades for new students in future courses, ranging from average course grades, over Random Forest (RF) regressors, to FMs. They found that FMs performed the best when not including any other information than the student-course interaction, while with additional information, FMs and Random Forest regressors performed similarly well. 
We are inspired by this work and the ability of FMs to capture high-cardinality data like user-item-context interactions. Consequently, we adopt a similar strategy and compare three different approaches to modeling player difficulty, including RF and FM. 

\section{Case study: Lily's Garden}

\begin{figure}[t]
    \begin{subfigure}{1\columnwidth}
    \includegraphics[width=1\columnwidth]{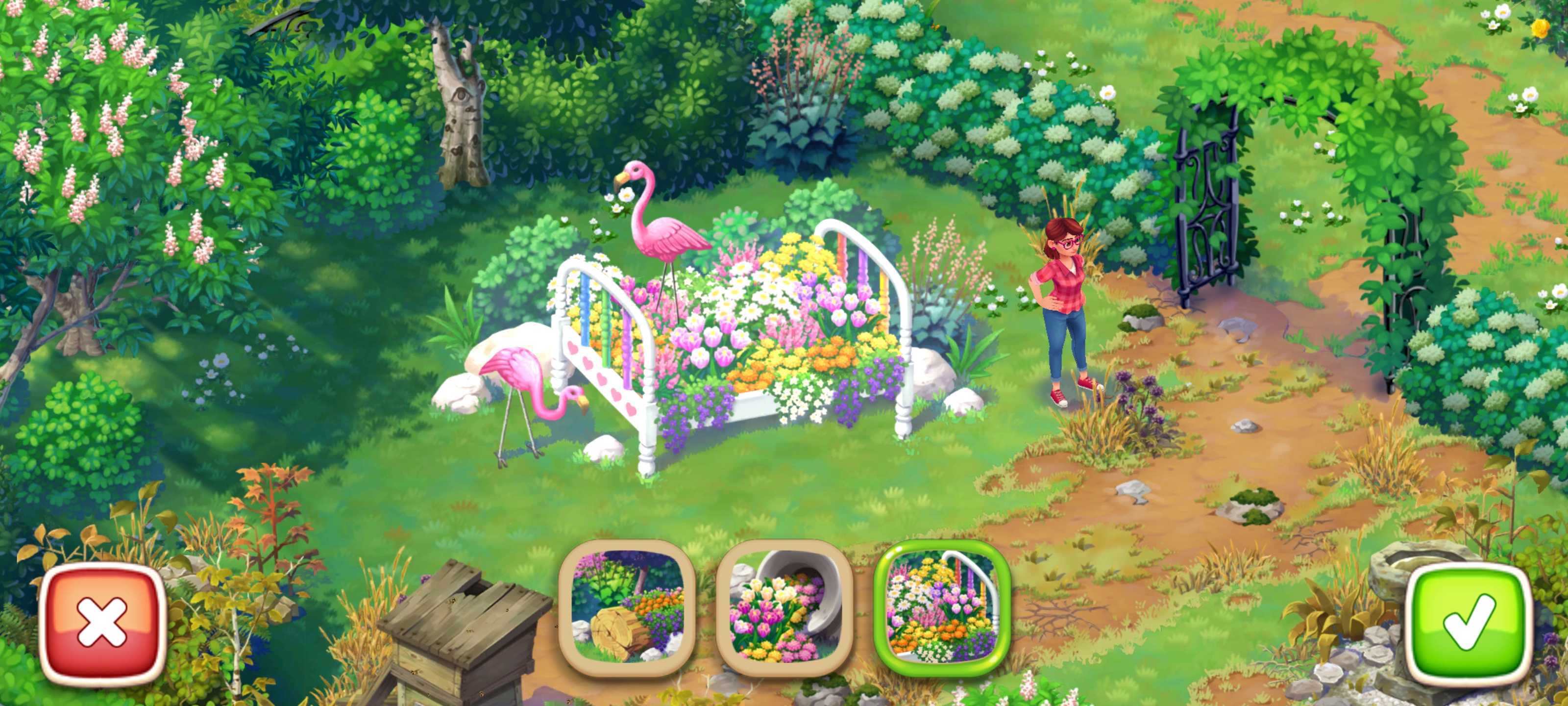}
    \caption{Metagame garden scene, with Lily being offered three flowerbed decoration options. Realizing an option and progressing in the storyline requires points, which the player collects by solving puzzles.}
    \end{subfigure}
    \begin{subfigure}{1\columnwidth}
    \includegraphics[width=1\columnwidth]{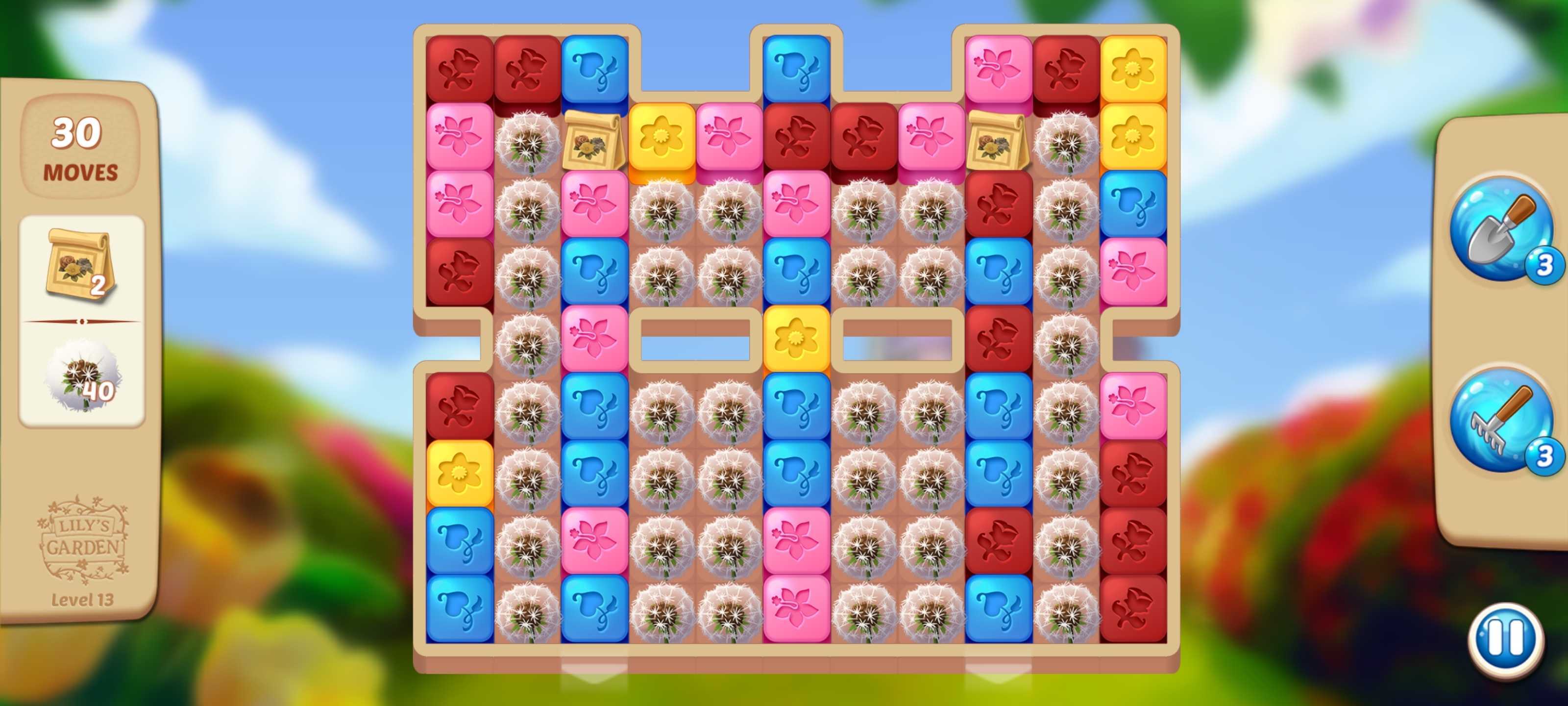}
    \caption{Example of a puzzle level. The goal is to collect certain board pieces, shown on the left side, within a limited number of moves.}
    \end{subfigure}
    \caption{Lily's Garden: interplay of meta and puzzle game.}
    \label{fig:lilys garden}
    \Description[Two screenshots of the main game]{Figure 1
Two screenshots from Lily's Garden, the game used for our case study. The purpose of the screenshots is to show the two main gameplay components of the game: the narrative metagame and the puzzle game within. 

Figure 1.a
Screenshot of Lily's Garden's narrative metagame where the player must fix and decorate the garden. In the example shown here, the player has three options for changing the flowerbed decoration piece in the garden.

Figure 1.b
Screenshot of level 13 from Lily's Garden's blast-type puzzle game. In the puzzle game, the player must collect certain board pieces within a limited number of moves to gain more credits for garden work and to progress in the narrative. The screenshot shows the gameboard is in the middle. On the left side, it shows the number of remaining moves to complete the level, and the specific goal pieces to be collected: two seed bags and 40 daffodils. The right side shows the available boosters actions which the player can use to complete the level.}
\end{figure}

We study the use of FMs for predicting individual player difficulty in the commercial game Lily's Garden by Tactile Games. Released in early 2019, Lily's Garden is a casual mobile puzzle game with more than 6,000 levels and one million daily active users worldwide.

The gameplay has two main components (Fig. \ref{fig:lilys garden}). In a  narrative-driven meta game, the player is confronted with an abandoned garden in which they can unlock new areas, make decorative choices and progress in the story by spending points. These points can be acquired by solving successive puzzle game levels unlocked at specific times in the storyline. This study focuses on predicting individual player difficulty for these puzzle levels. 

To complete a puzzle, the player must collect specific goal pieces on the board within a given move limit. The core gameplay consists of tapping on board piece clusters to clear them, destroy adjacent pieces, and hereby collect the goal pieces. By tapping on clusters with more than five, eight or ten pieces, the player can create power pieces capable of clearing large parts of the board. In some levels, forging such power pieces is strictly necessary to succeed, and more advanced strategies involve their combination for enhanced effects.

The game implements a free-to-play model. The player can use in-game and real currency to buy help in the form of power pieces and other boosters. Additionally, there are certain game events that provide such boosters for free. Moreover, players can purchase an additional five moves for the current attempt if they fail to complete all the goals within the initial move limit. While purchases are the main monetization avenues of the game, the designers ensure that every level can be completed without bought assistance.

\begin{figure}[t]
    \centering
    \includegraphics[width=1\columnwidth]{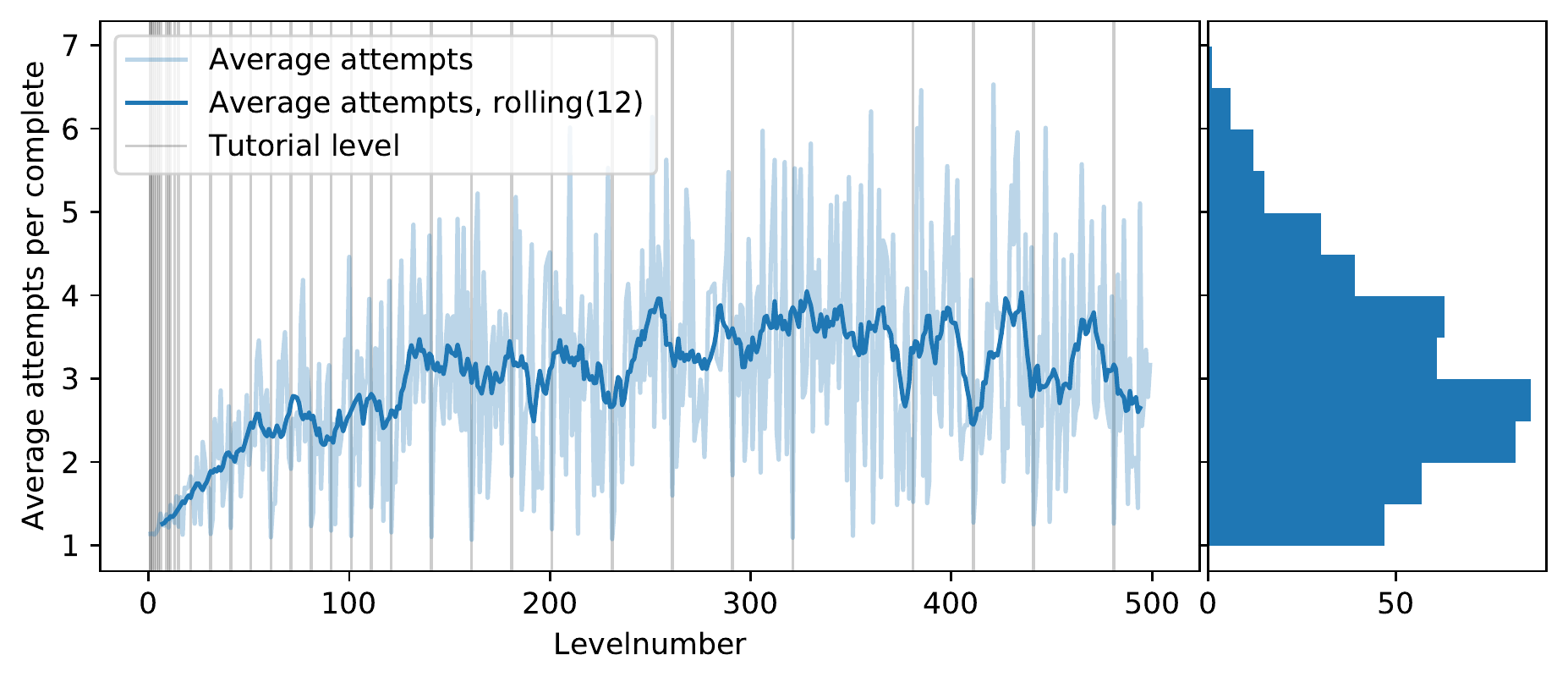}
    \caption{The average number of attempts per level completion for the first 500 levels. The difficulty trend is illustrated as a moving mean with a centered window of size 12 -- twice the length of typical designed level sequences.}
    \label{fig:leveldifficultycurve}
    \Description[A line graph that shows the difficulty over the first 500 levels]{Figure 2
Line graph showing the average difficulty for each of the first 500 levels. The x-axis shows the level number in ascending order, and the y-axis the average number of attempts per complete as our operationalization of difficulty. A rolling mean with window size 12 is also plotted to show the trend. Grey bars are overlaid to distinguish tutorial levels within the level sequence. The line starts at ca. 1.4 attempts on average for the earliest levels and rises with strong fluctuations to about 3 attempts on average for later levels. The line graph is complemented on the right with a histogram of the average number of attempts over all levels. It shows that most players require between 2.5 and 3 attempts per level complete.}
\end{figure}

For our model, we adopt the existing operationalization of level difficulty as the average number of attempts that players require to complete a level. An analysis of player data from Lily's Garden (Sec.~\ref{sec:data} provides an overview of this data) shows the average number of attempts over the whole level range (Figure \ref{fig:leveldifficultycurve}). The first few levels ($<10$) contain multiple tutorial levels where the gameplay is streamlined and players are restricted to certain moves.
After these levels with almost guaranteed wins, the difficulty slowly ramps up -- a common design pattern to engage players early on \cite{linehan_learning_2014}. 

To understand what level design aspects can affect (intrinsic) difficulty, we interviewed the team of level designers of Lily's Garden and identified a number of candidate level features that could be relevant for predicting difficulty.
This includes quantifiable features such as the move limit (higher limits lead to easier levels \cite{kristensen2021statistical}), the number of goals, and the entropy of the pieces' color distribution (the closer to uniform, the harder the level due to power pieces being harder to create).
Other features, such as the level layout, board piece complexity, or reliance on power pieces, are harder to quantify but will still affect the difficulty in non-trivial ways. Defining descriptors that can fully encapsulate these intricacies is therefore rather challenging and can lead to less expressive and accurate difficulty prediction models.

\begin{figure}[t]
    \centering
    \includegraphics[trim={0 0 0 0.68cm},clip,width=1\columnwidth]{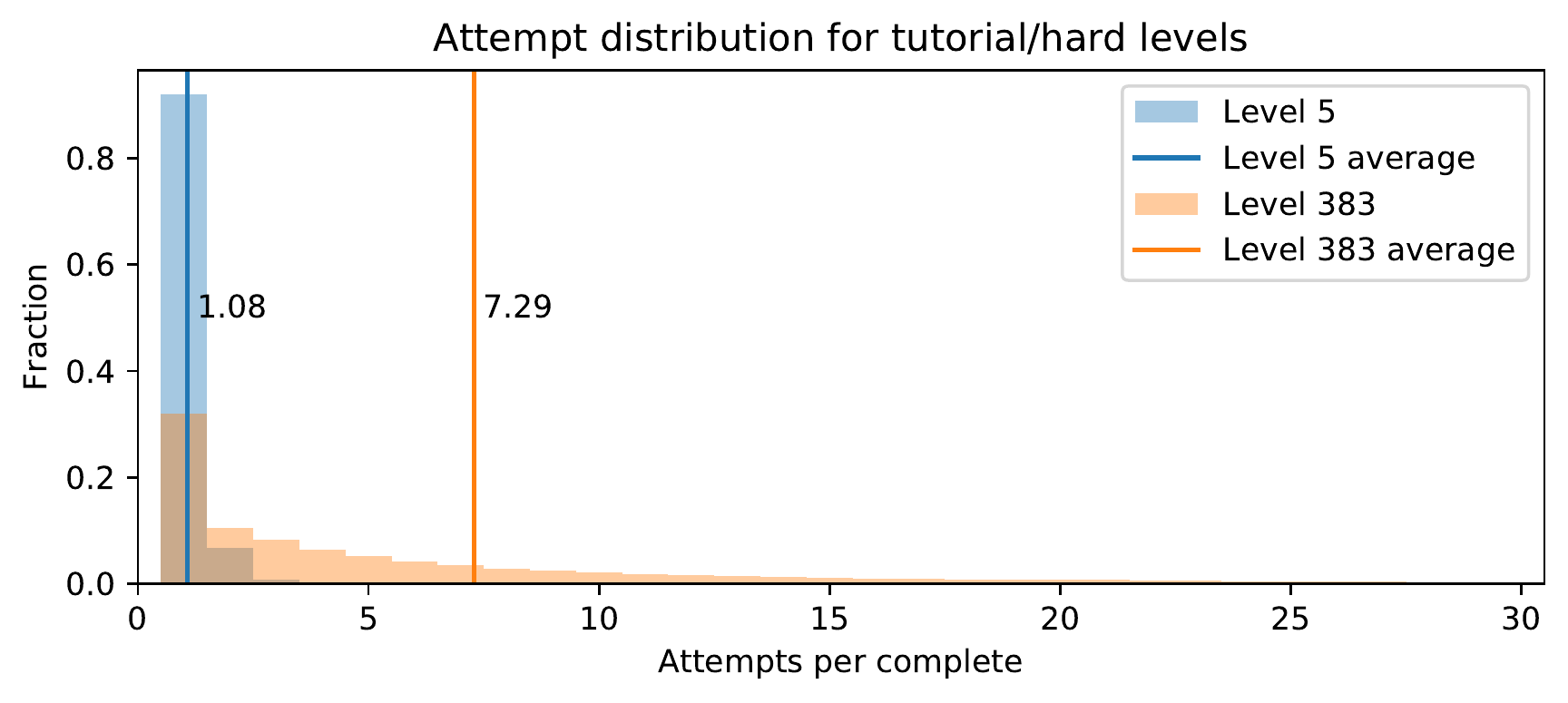}
    \caption{Comparison of player attempt distributions for a tutorial (level 5) and a hard level (level 383).}
    \label{fig:example attempt distribution}
    \Description[A historgram of the attempt distribution of a hard and easy level.]{Figure 3
Two overlaid histograms of the attempts per level complete on an easy and a hard level (level 5 and 383, respectively). Two vertical lines are also plotted to show the average number of attempts on the levels. The mode of both distributions is one, but the variance in attempts for the harder level 383 is much larger and can extend to more than 30 attempts per complete.}
\end{figure}

In addition to the strong fluctuations in the average number of attempts over the level range, a more detailed analysis also highlights large differences between individual completion rates both with respect to the players and the levels (Fig. \ref{fig:example attempt distribution}). In early, less (intrinsically) difficult levels with fewer than 1.5 attempts per complete on average, most people (>90\%) require 1 attempt, with the remaining players requiring a little more. For later levels with greater (intrinsic) difficulty, we find a large variance in attempts and a long tail distribution extending to more than 30 attempts per complete. 
These individual differences, paired with strong fluctuations in average difficulty, make Lily's Garden an ideal, challenging candidate for studying personalized player difficulty prediction.

\section{Methods}
\label{sec:methods}
In accordance with existing work on puzzle games (Sec.~\ref{sec:operationalisation}), we operationalize difficulty by the number of attempts a player will spend on the level.
We consequently frame our individual player difficulty prediction task as a regression problem, where the target is to estimate the number of attempts a specific player will spend on a specific level.
For this purpose, we compare four different methods:

\begin{itemize}
    \item Naive baseline (\textbf{NB}): Average attempts by other players.
    \item Random Forest regression (\textbf{RF}): Ensemble prediction from multiple decision trees that utilize aggregated player behavior data over the observed levels.
    \item Factorization Machines (\textbf{FM}): A general regression model that uses a feature embedding to describe interactions between variables (e.g.~user-item interactions).
    \item Factorization Machines with Relational Data (\textbf{FM+feat}): As \textbf{FM}, but also includes the descriptive variables used in the RF method (e.g. user-item-feature interactions).
\end{itemize}

To answer our first research question (Sec.~\ref{sec:intro}), we compare these models based on their prediction error. We moreover analyze how this error changes based on the number of levels that we observed the players for. In other words, we identify the number of required observations to push the error below a certain threshold and thus answer our second research question.



\subsection{Naive Baseline}
Given that much related work focuses on predicting difficulty for a player population rather than individuals, we chose the player-average number of attempts per level complete as our naive prediction baseline. We calculate this non-personalized prediction on the players' data from our training set as illustrated in Fig. \ref{fig:leveldifficultycurve} using a linear regression model,
\begin{equation}
    \hat{y} = w_0 + \sum_{\ell=1}^L{w_\ell x_\ell},
    \label{eq:basic linear equation}
\end{equation}
where $w_0=0$, $w_\ell$ is the attempts on level $\ell$ averaged over all other players, L is the total number of levels, and $x_\ell \in \{0, 1\}$.

This non-personalized baseline is what game designers currently use in practice for estimating level difficulty. Hence, any improvements over this baseline can directly inform game designers of the compared methods' benefits.

\subsection{Random Forest Regression}
\label{sec:rf model}
As mentioned in Sec.~\ref{sec:related_work_models}, a Random Forest (RF) regression model has previously been shown to deliver comparable performance to FMs \cite{sweeney_next-term_2016} in a related task. Consequently, we train an RF regressor on player and level features in our comparison. 

RF is an ensemble method based on multiple random trees, i.e., decision trees
$h(\mathbf{x; \theta_t}), t = 1,..., T; \mathbf{\theta_t}$ with \textit{i.d.d.} random vectors, 
where the nodes of each tree are split using a random set of features and subsets of the data.
For regression problems, the split is decided based on which feature leads to the largest decrease in the absolute or squared error. The final prediction then combines the predictions from the trees into an average prediction, $y = \hat{h}(\mathbf{x})$.
This ensemble approach enables modeling more complex non-linear behavior and is less likely to overfit compared to a single random tree.


We use the Random Forest regressor implementation from the \emph{scikit-learn} library (version 1.0.2) \cite{scikit-learn} 
with the following hyperparameters and settings: \verb|n_estimators=150|, \verb|max_depth=None|, 

\noindent\verb|min_samples_split=2|, \verb|min_weight_fraction=0.0|, 

\noindent\verb|max_features="auto" [=n_features]|, \verb|max_leaf_nodes=None|, \verb|min_impurity_decrease=0.0|. Due to the size of our data, we employ an incremental training method\footnote{\url{https://github.com/garethjns/IncrementalTrees}}.

\subsection{Factorization Machine}
Factorization machines (FMs) are a class of factorization models that can be used as a general predictor for classification, regression, and ranking tasks \cite{rendle2010factorization}. 
They are similar to linear regression models but instead model second-order terms as an interaction between variables using a feature embedding:

\begin{equation}
    \hat{y} = w_0 + \sum_{i=1}^n{w_i x_i} + \sum_{i=1}^n\sum_{j=i+1}^n{\langle \mathbf{v_i, v_j} \rangle x_i x_j},
    \label{eq:fm equation}
\end{equation}

where $w_0$ is the 0th order term or global bias, $w_i$ is the first-order term and describes the bias of the $i$'th variable, and $\mathbf{v_i}$ is a second-order feature embedding vector of the $i$'th variable. $\langle \mathbf{v_i, v_j} \rangle$ describes the interaction between two variables as the dot product:

\begin{equation*}
    \langle \mathbf{v_i, v_j} \rangle = \sum_{f=1}^k{v_{i, f} \cdot v_{j,f}},
\end{equation*}

where $k$ is the number of latent factors and a hyperparameter that must be chosen.

This learned embedding is what enables modeling unseen interactions in the data, which also makes FMs widely used for recommendation systems, including games recommendations \cite{anwar_game_2017,cheuque_recommender_2019}, where user-item interactions are typically very sparse. 

The input data is not restricted to the rating-user-item format. 
It is possible to use other contextual information \cite{rendle2013scaling}, including
\begin{itemize}
    \item One-hot encoding of previous user interactions on other items.
    \item User and item descriptors (both numeric and categorical, e.g. age or labels).
    \item Implicit feedback data.
\end{itemize}

Adding additional features typically increases the data set complexity by $n \times f$, where $n$ is the dataset length and $f$ is the number of additional features. However, using FMs with a \emph{relational data} block structure to make use of repeated patterns, as suggested by Rendle et al. \cite{rendle2013scaling}, can greatly reduce the computational complexity and make the method scale to very large datasets.
Additionally, Rendle et al. found that FMs with such relational data showed a consistent performance increase over FMs without relational data.

We use the original implementation of LibFM by Steffen Rendle \cite{rendle_factorization_2012}. We train each model for 1000 iterations using Markov Chain Monte-Carlo (MCMC) with an initial standard deviation of 1 to sample $v_i$ for FM models and $0.1$ for FM+feat models. Based on brief experiments and the nature of the attempt distribution data, we do not include a global bias term $w_0$.

\section{Data}
\label{sec:data}

\begin{table*}[ht]
\begin{tabular*}{0.9\textwidth}{llp{9.5cm}}
\textbf{Type of feature}                               & \textbf{Name} & \textbf{Description}                              \\ \hline
\multicolumn{1}{c|}{\multirow{10}{*}{Player features}} & Attempts & Number of attempts on levels \\
\multicolumn{1}{c|}{}                                  & Moves used  & Number of moves used relative to the move limit when completing the level \\
\multicolumn{1}{c|}{}                                  & Pre-game boosters & Boosters that can be used before starting the level \\
\multicolumn{1}{c|}{}                                  & In-game boosters  & Boosters that can be used while playing the level \\
\multicolumn{1}{c|}{}                                  & Powerpieces, total & Number of power pieces created while playing the level \\
\multicolumn{1}{c|}{}                                  & Powerpieces, combos & Number of power piece combinations created while playing the level\\
\multicolumn{1}{c|}{}                                  & Rockets, solo & Number of rockets created and used on their own\\
\multicolumn{1}{c|}{}                                  & Rocket-bomb combo & Number of rocket-bomb combos created and used \\
\multicolumn{1}{c|}{}                                  & Rocket-magic combo & Number of rocket-magic combos created and used \\
\multicolumn{1}{c|}{}                                  & Bomb-magic combo & Number of bomb-magic combos created and used \\ \hline
\multicolumn{1}{c|}{\multirow{9}{*}{Level attributes}} & Attempts & Average number of attempts on level by players in training set \\
\multicolumn{1}{c|}{}                                  & Color entropy & Entropy of color spawning weights; $S=-\sum_{i}{p_i\log p_i}$ \\
\multicolumn{1}{c|}{}                                  & Colors & Number of unique colors in the level \\
\multicolumn{1}{c|}{}                                  & SpreadingBlocker, cg & Levels with a spreading blocker as collect goal (cg) \\
\multicolumn{1}{c|}{}                                  & LayerCake, cg & Levels with a specific blocker with 3 hitpoints as a collect goal \\
\multicolumn{1}{c|}{}                                  & ConsecutiveBlocker, cg & Levels with a blocker that requires two attacks in a row as a collect goal \\
\multicolumn{1}{c|}{}                                  & MegaMultiColorBlocker & Levels with a large blocker that requires matching multiple colors to remove \\
\multicolumn{1}{c|}{}                                  & Teleport & Levels with a teleport mechanic that transports pieces around the board \\ 
\hline
\end{tabular*}
\parbox{1.8\columnwidth}{\caption{Features investigated for RF and FM+feat. The player features are aggregated means on the first $n$ observed levels.}\label{tab: supervised features}}
\Description[Table containing an overview of the different data types.]{Table 1
Table of the player and level features used for the random forest and factorization machines models. The table shows the specific features and their descriptions, categorized into player features and level attributes. The player features are individually aggregated over all observed levels. Amongst others, they comprise the average number of attempts, average number of moves used and average number of power pieces created. The level features include, amongst others, the average number of attempts on the level by all players in the training set, information about the weight distribution of spawning colors, and information on the presence of specific board pieces that require special strategies.}
\end{table*}

\begin{figure}[ht]
    \centering
    \includegraphics[width=0.9\columnwidth]{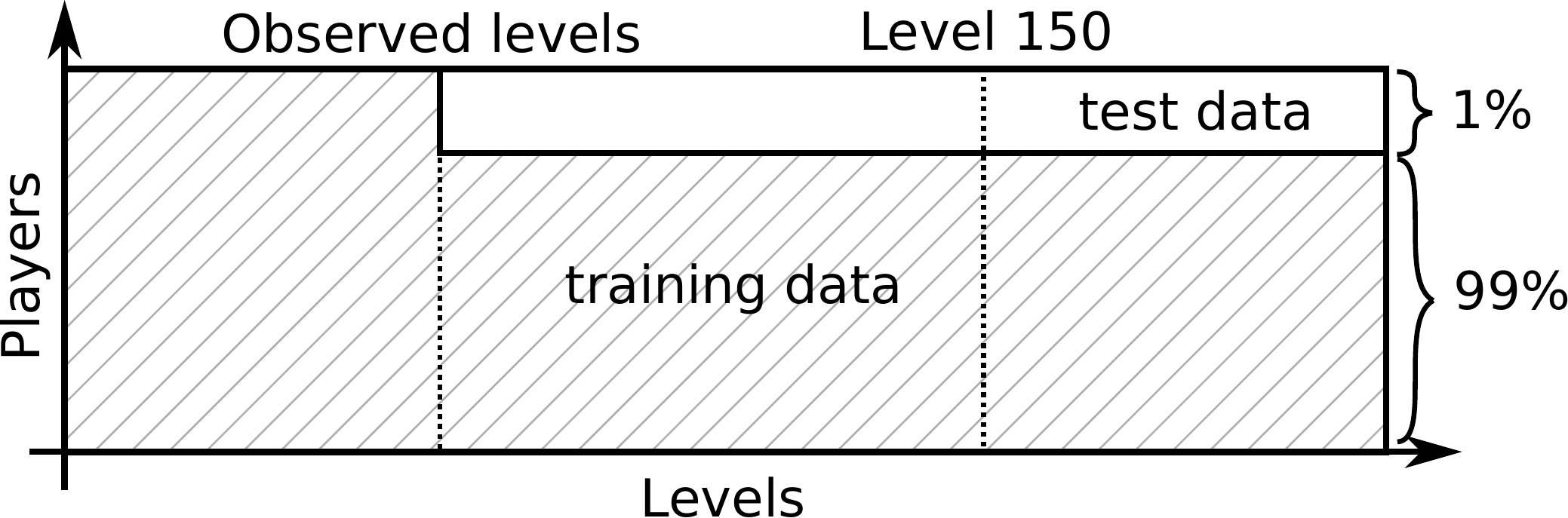}
    \caption{Train/test data split illustration. All data of 99\% of players is used for training. Additionally, the training utilizes initial observations (``Observed levels'') from the 1\% of players who constitute the test set.}
    \label{fig:fm data}
    \Description[A diagram of how the data is split in our experiments]{Figure 4
Diagram illustrating how we split our data for training and testing the factorization machine model. It shows that data from 99\% of players across all levels is used for training and that the remaining 1\% is used for testing. The model performance is evaluated on the data of the players in the test set after level 150. The remaining test set data, before level 150, is included in different amounts as observations to update the model. The goal of this is to understand how many observations of a player are necessary to discern them from the average player.}
\end{figure}

The data used in this study was collected from 2021-06-01 to 2021-11-30 from the game Lily's Garden and consists of 759,382 players who have all, in this period, played the game beginning with the first level and at least up to level 200. In free-to-play mobile games, it is common to have a large churn rate at the beginning of the game from players that do not interact meaningfully with the game, so this condition ensures both, that the whole history of each player is complete, and that the included players share the same minimum engagement level. Due to the long tail of the attempts distribution (Fig. \ref{fig:example attempt distribution}), for numerical stability, we truncate attempts with more than 30 attempts to 30, which affects 0.34\% of the data.

We split the data to match a realistic use case: We select 1\% of the players to represent ``new'' players to test the methods on, and the remaining 99\% of the players are considered ``old'' players who have started playing earlier and have already progressed far in the game. We use this old player data for training all models. 

Additionally, as illustrated in Fig. \ref{fig:fm data}, FM training also utilizes some initial observations of the new players. This corresponds to the model being periodically updated to improve its predictions as new players progress through the game and more observations become available. It is also necessary for FM: without  observation of the given player during training, the model does not learn the bias and embedding of this player (cold start problem).

In our performance evaluations, we report the results with different numbers of initial observations. The results are always computed from levels after 150 to maintain a consistent test set even when the number of initial observations changes. 


The RF and FM+feat methods require feature vectors that describe the players and levels. These features were selected based on domain knowledge from level designers and are shown in Table \ref{tab: supervised features}. RF methods do not deal well with high-cardinality data, so it is not possible to one-hot encode players and levels in this case. Instead, the player features for all players are aggregated and averaged across the first $n$ observed levels. 
This means that any prediction of a player depends on their early performance in the game and does not take recent observations into account. Otherwise, players are not comparable through their features due to each player being at different stages of the game with different types of levels, difficulties, required strategies, etc. The level attributes are static and do not change depending on the number of observations.

\section{Results}
In the following, we first describe the prediction task to put the baseline prediction and error metric in context. We then identify how many observations are necessary to beat the baseline to answer our first and second research questions. 
Lastly, in order to answer our third research question, we provide an interpretation of the model parameters and identify key game levels for understanding the model and thus providing valuable insights for the level designers.

\subsection{Baseline Prediction and Error}

Both the RF and FM models have been optimized using root mean square error (RMSE).
However, the underlying distribution of attempts, as shown in Fig. \ref{fig:example attempt distribution}, follows a geometric-like distribution, where the most common value is 1, and especially hard levels exhibit a long tail that drives the average attempts up.
This long tail on hard levels can yield a large RMSE on said levels, leading to the hard levels having a large effect on the model optimization. To provide a more complete picture of the models' performance and support model comparisons on easier levels, we also report the mean absolute error (MAE) next to RMSE.

Crucially, we cannot expect this baseline nor any of our methods to yield a close-to-zero prediction error. This is because each outcome of a level playing attempt is also governed by aleatoric uncertainty, which the model cannot account for. Each prediction captures the expected value for a given player-level combination.

Calculating the baseline error by aggregating all data points, we find the prediction errors across the whole level range to be $\textup{RMSE}_\textup{all} = 3.86$ and $\textup{MAE}_\textup{all} = 2.33$. The errors after level 150 are $\textup{RMSE}_{\ell >150} = 4.10$ and $\textup{MAE}_{\ell>150} = 2.53$. The larger error on $\ell>150$ is due to these levels being generally (intrinsically) more difficult (Fig. \ref{fig:leveldifficultycurve}) and thus having a larger attempt variance.


\subsection{Effect of Observed Level Count}

Before being able to differentiate between players meaningfully, we require sufficiently many discernible observations of their gameplay. The first 10 levels introduce the core gameplay, and new mechanics are then introduced in every tenth tutorial level (21, 31, 41, ... see Fig. \ref{fig:leveldifficultycurve}). We, therefore, compare the methods when trained at 6 points of a player's progress: at 10, 20, 30, 50, 100 and 150 levels. To avoid information leaking between training and test sets, we evaluate the predictions on the test users after level 150.

\begin{figure}[t]
    \centering
    \includegraphics[trim={0 0 0 1cm},clip,width=1\columnwidth]{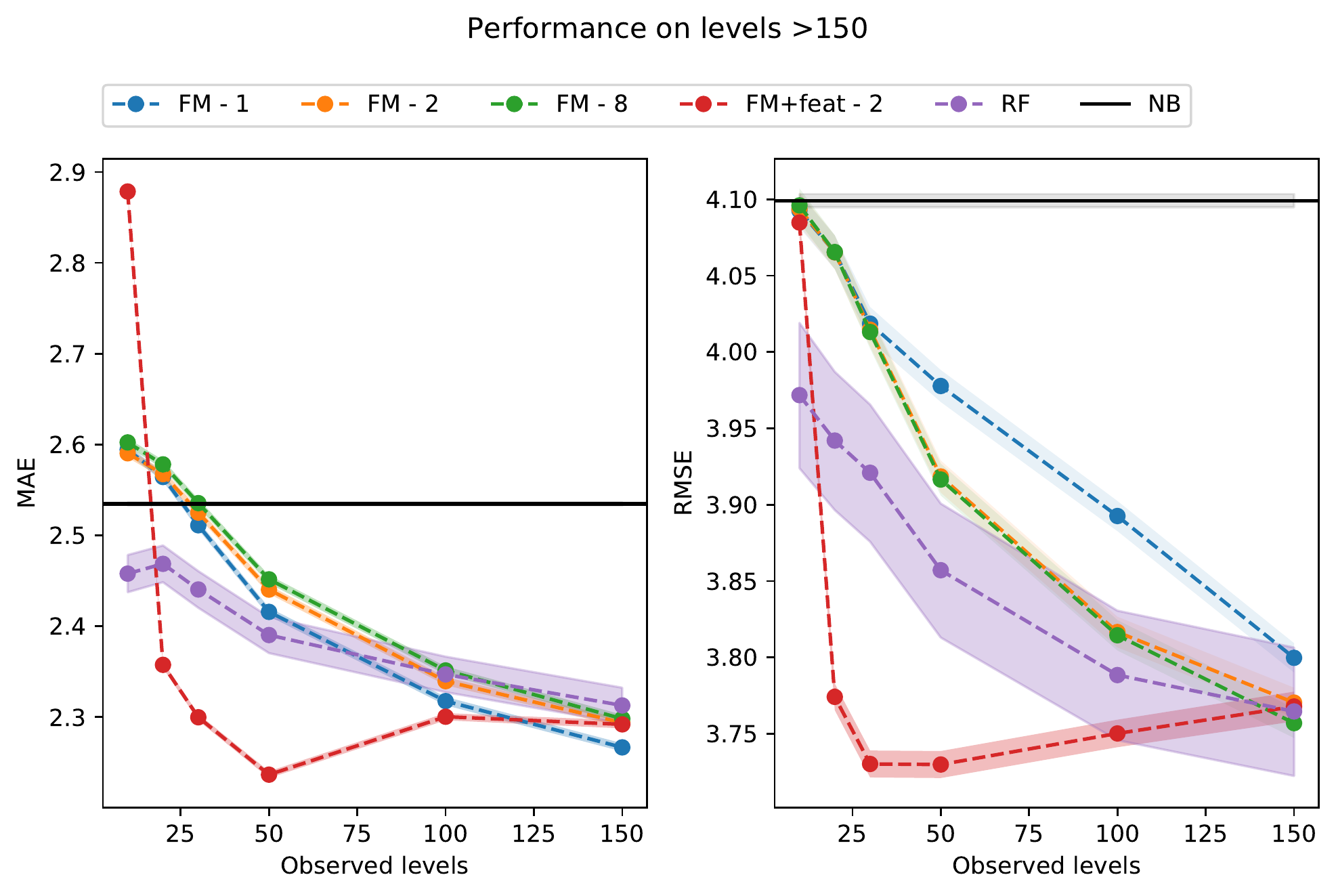}
    \caption{MAE and RMSE on test levels $\ell>150$ for different observed level counts. The shaded area shows the 95\% confidence interval around the means.}
    \label{fig:mae vs training levels}
    \Description[The mean absolute error and root mean squared error of the models.]{Figure 5
Two line graphs showing the model errors for increasing counts of observed levels. The x-axis shows the number of level observations, and the y-axis the mean absolute error (MAE) in the left-hand graph, and the root mean square error (RMSE) in the right-hand graph. 6 lines are shown. The black line representing the naive baseline is constant. The lines representing the random forest and standard factorization machine models decrease with increasing number of observations below the baseline. The random forest model performs better compared to the naive baseline after 10 observations. The factorization machines require between 20-30 observations before their MAE is less than the naive baseline, but only 10 observations before their RMSE is lower. All lines decrease continuously but the the red lines in both plots, representing the factorization machine model with additional features, which have minima between 30 and 50 observations. At this minimum, the model performs considerably better than the random forest.}
\end{figure}

\rebut{We tested our FMs with 1, 2, 4, 8, 16, and 32 factors, but we leave out the 4, 16, and 32-factor models in the presented results for clarity of visualization since they do not alter or further inform our conclusions.}
The MAE and RMSE of all tested models are shown in Fig.~\ref{fig:mae vs training levels}, along with the 95\% confidence intervals. 
The FM models without additional features (i.e., excluding FM+feat-2) all have similar performance, with the 1-factor model performing better in terms of MAE and the other FM models performing better in terms of RMSE. This suggests that using a single latent factor is not enough to capture the large variance in high-difficulty levels (see Fig. \ref{fig:example attempt distribution}), but it makes the model less likely to overfit and perform worse on easier levels. The RMSE plot shows that these FM models are on par with the baseline prediction for as little as 10 observed levels, and, as more levels are observed, the prediction further improves over the baseline.

The predictions can be further improved while requiring fewer observations by including additional features as described in Table \ref{tab: supervised features}. This holds for both the RF and FM+feat models. The RM+feat-2 model shows superior performance, but its error increases after 50 observed levels. We consider this an effect of overfitting to the additional data since the training error for all the FM+feat models appears to be around $\textup{RMSE}_\textup{train} = 3.2$. 

This highlights the importance of feature engineering, and while more sophisticated features could be utilized, the basic FMs presented here are game-agnostic and scale easily, providing a feasible method for game designers to employ. Our results also show that, even when additional information is available and included as features in the RF model, FMs are still able to extract more relevant information from the player-level-feature interactions.

\begin{figure}[t]
    \centering
    \includegraphics[trim={0 0 0 0.67cm},clip,width=1\columnwidth]{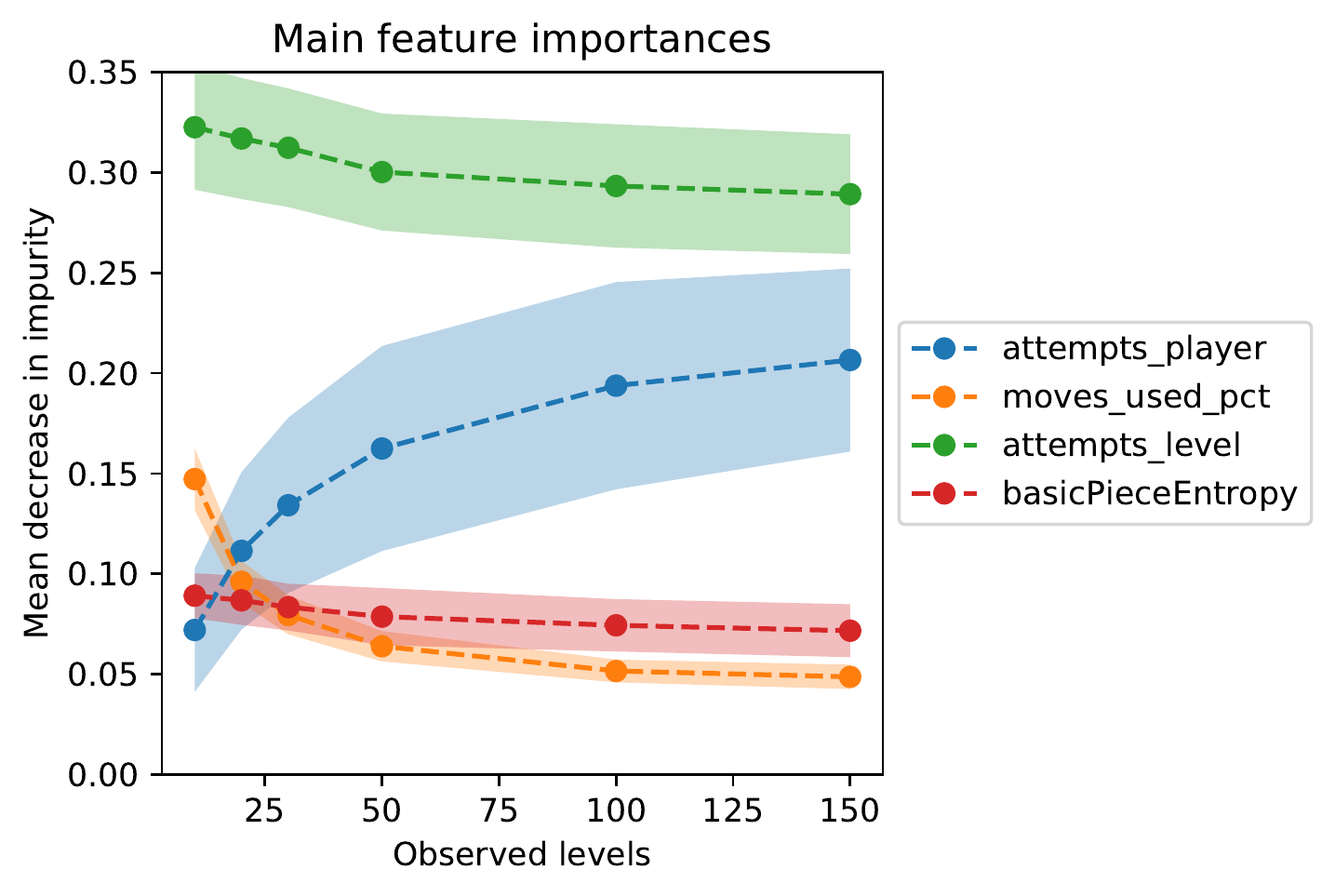}
    \caption{Mean feature importances and one standard deviation for the Random Forest (RF) regressor. Only features with a relative significance above 0.05 are shown.}
    \label{fig:feature importances}
    \Description[Random forest feature importances as a function of the number of observed levels.]{Figure 6
Line graph showing the random forest feature importance for increasing counts of of observed levels. The x-axis shows the number of level observations, and the y-axis the mean decrease in impurity. The 4 lines represent the most important features. The feature describing the average attempts on a level is the most important feature across all data points. The significance of the average number of attempts by the player become more important as more observations are included. The feature describing the average number of moves decreases as more observations are included. The feature describing the color entropy of the level decreases slightly as more observations are included.}
\end{figure}

To explore why early predictions can be improved by including additional data comprising more fine-grained descriptions of players, we analyze the feature importances from the RF model.
Fig. \ref{fig:feature importances} shows how the four most important features change depending on the level observation count. We find that the model utilizes additional information other than the number of attempts: early on, more fine-grained behavior data such as the average number of moves is more important compared to later on, where the average number of player attempts becomes increasingly more important. However, with more observed levels, all models reach a similar degree of performance. While the FM+feat-2 model 
performance appears to stagnate, the FM approaches reach similar, if not better, performance. This stagnation may be caused by the feature aggregation, where the possible level of detail and stand-out behaviors are more washed out with an increasing number of observations. Other features, such as power piece and booster usage, all have a relative importance of around $0.05$, i.e., they are not uninformative. However, they are strongly correlated, which may reduce their joint importance, as mentioned in Sec. \ref{sec:rf model}.

\begin{figure}[b]
    \centering
    \includegraphics[width=1\columnwidth]{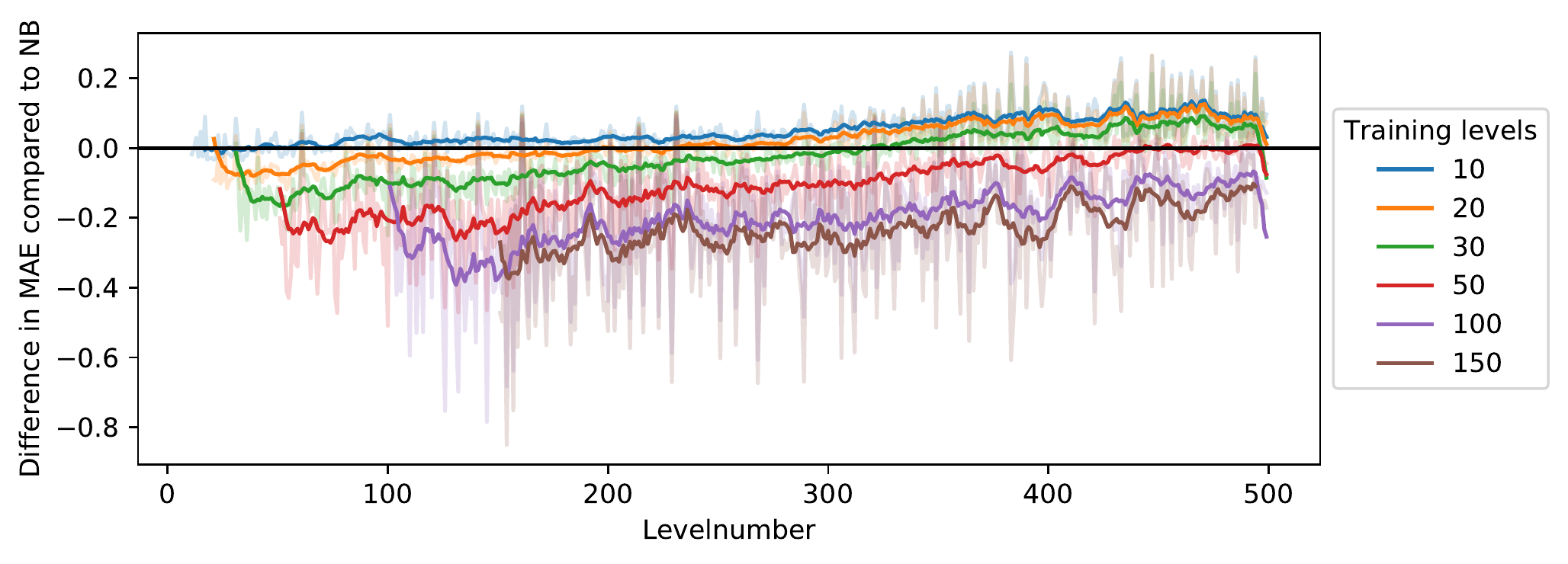}
    \caption{MAE difference between the prediction and baseline for the FM-2 model on the test user predictions for different observed level counts. The lines show the rolling average over 12 levels. Similar behavior was observed for the other models but is not shown for clarity.}
    \label{fig:baseline improvement nextlevels}
    \Description[Change in mean absolute error over the training levels.]{Figure 7
Line graph showing the difference in mean absolute error between a factorization machine model and the naive baseline across the whole level range. The x-axis shows the level number in ascending order, and the y-axis shows the error difference. 6 slightly opaque lines are shown for the models with 10, 20, 30, 50, 100 and 150 observations. The rolling means with window size 12 are also plotted for each model to show the trend. The more observations are included, the larger the difference between the factorization machine model and naive baseline becomes. For the models with 20 and 30 observations, the factorization machine predictions become worse than the baseline prediction after levels 200 and 300. The factorization machine models with more than 50 observations perform better, i.e. maintain a negative difference, than the baseline prediction for all levels. The difference becomes smaller with increasing level number.}
\end{figure}

We also analyzed how the accuracy of predictions varies depending on the specific level they are computed for. To this end, we plotted the MAE of the FM-2 model relative to the naive baseline, as shown in Fig. \ref{fig:baseline improvement nextlevels}. We omit a similar plot for the other models as they exhibited similar behavior.
We find that the predictions immediately after the last observed level are more accurate, with performance  deteriorating compared to the baseline on later levels. 
We conclude that, firstly, the FM prediction accuracy drops at later levels, and, secondly, more observed levels lead to a slower decline in predictive performance, meaning a higher observed level horizon allows a model to be used farther into the future.

The results presented in this section allow us to answer the first two research questions: Firstly, the basic FM model without additional information fares worse than the RF model with additional data until after 100 observed levels, where they converge to similar performance. However, the FM with the same additional data as the RF outperforms all other models after 20 observed levels.
Secondly, the number of observations that are necessary to discern a player from the average players depends on how detailed the available information is: with only the aggregate number of attempts, the FM method requires between 10 and 30, while more fine-grained data can enable predictions after 10 observations for all approaches.



\subsection{Interpreting FM Model Parameters}

To answer the third research question, we investigate the found FM model parameters in detail.
As already mentioned, the FM models with two or more factors seem to capture other aspects than the 1-factor model, and the FM-2 method yields comparable performance as the RF method after 100 observed levels. We thus limit the analysis to the 2-factor model with 100 observed levels.

\begin{figure}[t]
    \centering
    \includegraphics[width=1\columnwidth]{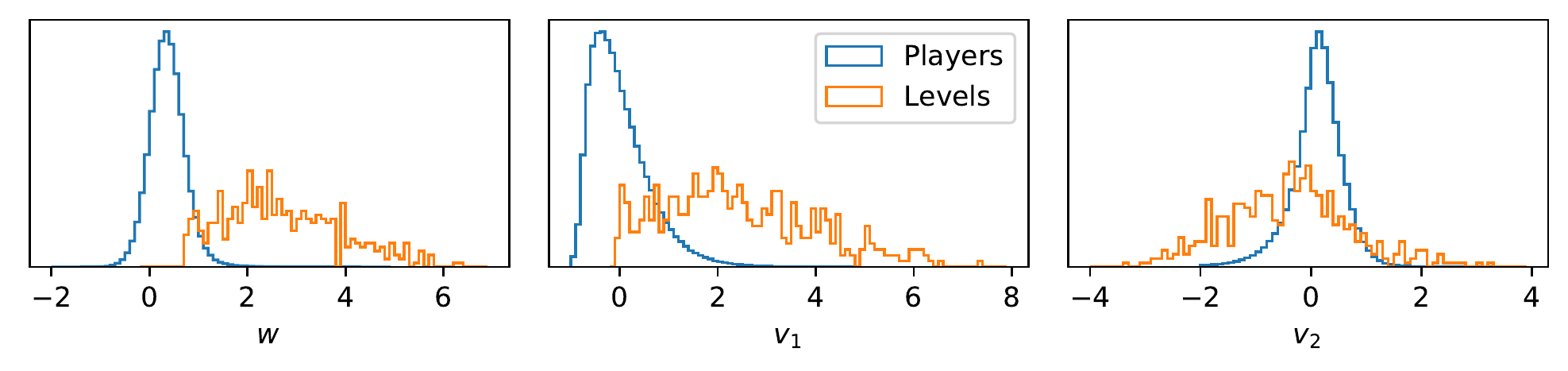}
    \caption{Histogram of the FM model parameters (100 observed levels, 2 latent factors).}
    \label{fig:model parameter histogram}
    \Description[Factorization Machine model parameter histograms.]{Figure 8
3 separate charts showing histograms of the factorization machine parameters. Each chart shows 2 overlaid histograms of the model parameters for the players and for the levels. The left chart shows the bias, w, the middle chart the first latent factor, v1, and the right chart the second latent factor, v2. The histogram of the bias, w, of the players follows a normal-like distribution and is in the range between -0.5 to 1. The histogram of the bias, w, of the levels is in the range between 1 to 6. The histogram of the first latent factor, v1, of the players follows a right-skewed normal-like distribution and is in the range between -1 to 2. The histogram of the first latent factor, v1, of the levels is almost strictly positive and is in the range between 0 to 7. The histogram of the second latent factor, v2, of the players follows a slightly left-skewed normal-like distribution and is in the range between -2 to 1. The histogram of the second latent factor, v2, of the levels follows a normal-like distribution and is in the range between -3 to 2.}
\end{figure}

FMs use two main parameters (Sec.~\ref{sec:methods}): the variable biases, $w$, and the latent factors that describe second-order interactions, $\mathbf{v}$ (ignoring subscripts for specific users/items/attributes).
As a first step, we consider histograms of the model parameters in Fig. \ref{fig:model parameter histogram} and separate the distributions by player and level parameters. We find that the distributions are distinctly different between players and levels for all three parameters. We consequently differentiate between levels and players in the following examination of each parameter before discussing their relationship to the player-level interactions.

\begin{figure}[h]
    \centering
    \includegraphics[width=0.9\columnwidth]{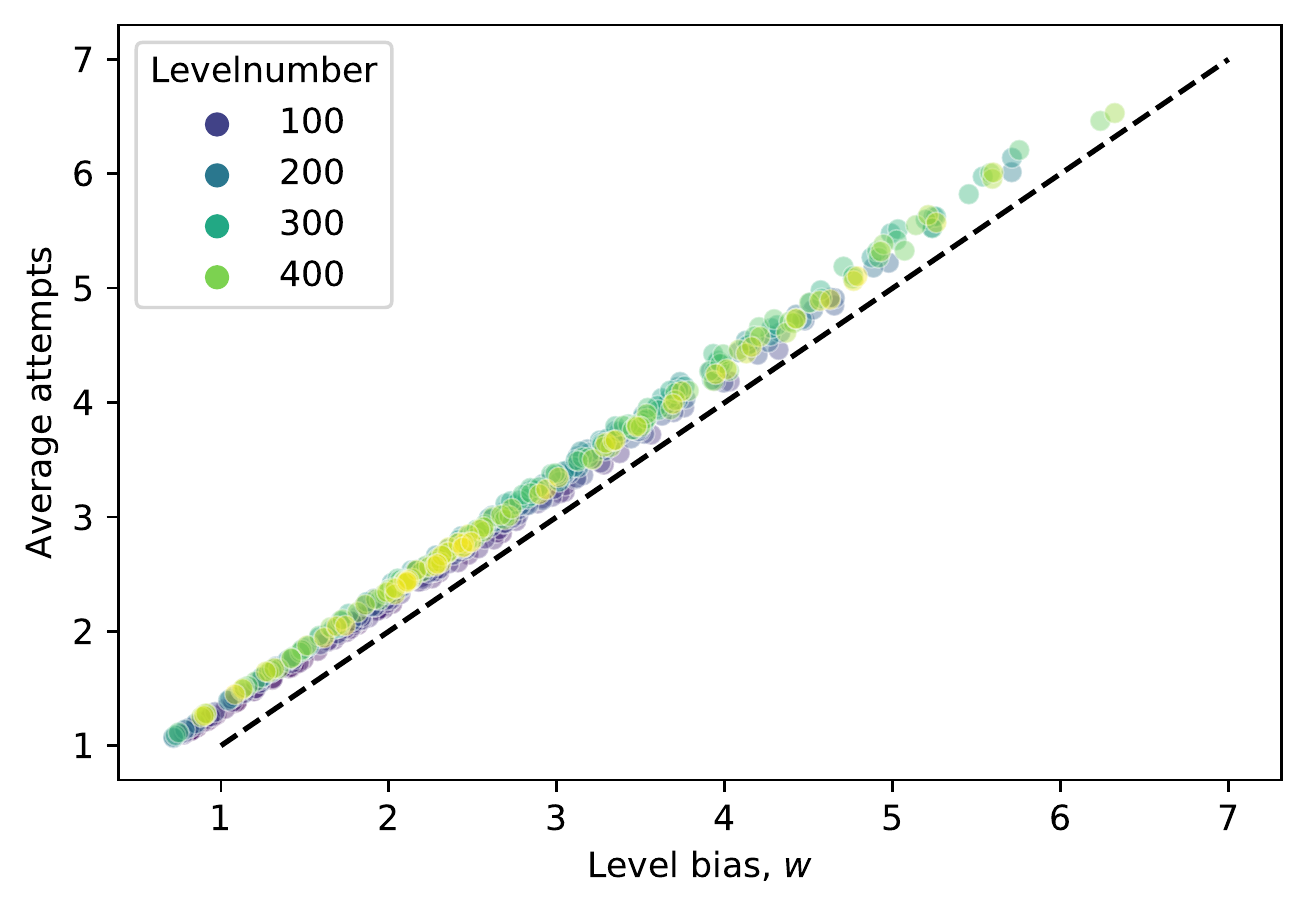}
    \caption{The average attempts on the level over the level bias, $w$. The black dashed line is the diagonal $w=\textup{attempts}$.}
    \label{fig:attempts vs levelbias}
    \Description[Scatterplot of level bias compared to actual attempts.]{Figure 9
Scatterplot showing the correlation between the average number of attempts on a level and the level biases, w. The x-axis shows the level bias, w, and the y-axis the average number of attempts per complete for the level. The colors represent the level number. A black diagonal lines is shown to indicate when the two values are equal. The dots are arranged on slightly offset diagonal. This highlights that the level bias, w, is strongly correlated with the average number of attempts per complete and that it is generally 0.5 lower than the average attempts per complete.}
\end{figure}

\textbf{Variable bias, $\mathbf{w}$:} For the levels, this variable is strongly correlated ($\rho_{\textup{Spearman}}=0.99$) with the average number of attempts on that level (Fig.~\ref{fig:attempts vs levelbias}). However, the $w$ parameter for the levels is not sufficient to match the underlying distribution of attempts, which is not surprising since that would correspond to the baseline model. 
As for the players, $w$ does not correlate strongly with any player-related metrics.
This suggests that the main first-order term comes from the levels, supporting an interpretation of $w$ as a baseline level difficulty.
The remaining variance must therefore be explained by the second-order interaction terms between player and level.

\textbf{First latent factor, $\mathbf{v_1}$:} For both the levels and players, this variable is also strongly correlated with the average number of attempts ($\rho_{\textup{Spearman}}=0.98$ and $0.70$, respectively), as well as the variance of the attempts ($\rho_{\textup{Spearman}}=0.99$ and $0.73$, respectively). We note that for the levels, $v_1$ is strictly positive, with the exception of 5 tutorial levels which are only slightly negative. This suggests that $v_1$ captures a variance effect for each level, where the sign of the interaction depends solely on the player. The amplitude then reflects the consistency of the level/player: a value close to 0 for the level signals little variation between the players (it may only require one simple strategy to win, e.g.~in tutorials), while a large $v_1$ suggests a large variation (e.g.~it may require multiple strategies, or come with high aleatoric uncertainty). For the player, $v_1$ is more reflective of their skill level: for $v_1<0$, we expect a player to use consistently fewer attempts than their peers, especially on harder levels where skill and strategy matter more. Conversely, $v_1>0$ indicates that the player struggles to employ winning strategies.

\begin{figure}[t]
    \centering
    \includegraphics[width=0.95\columnwidth]{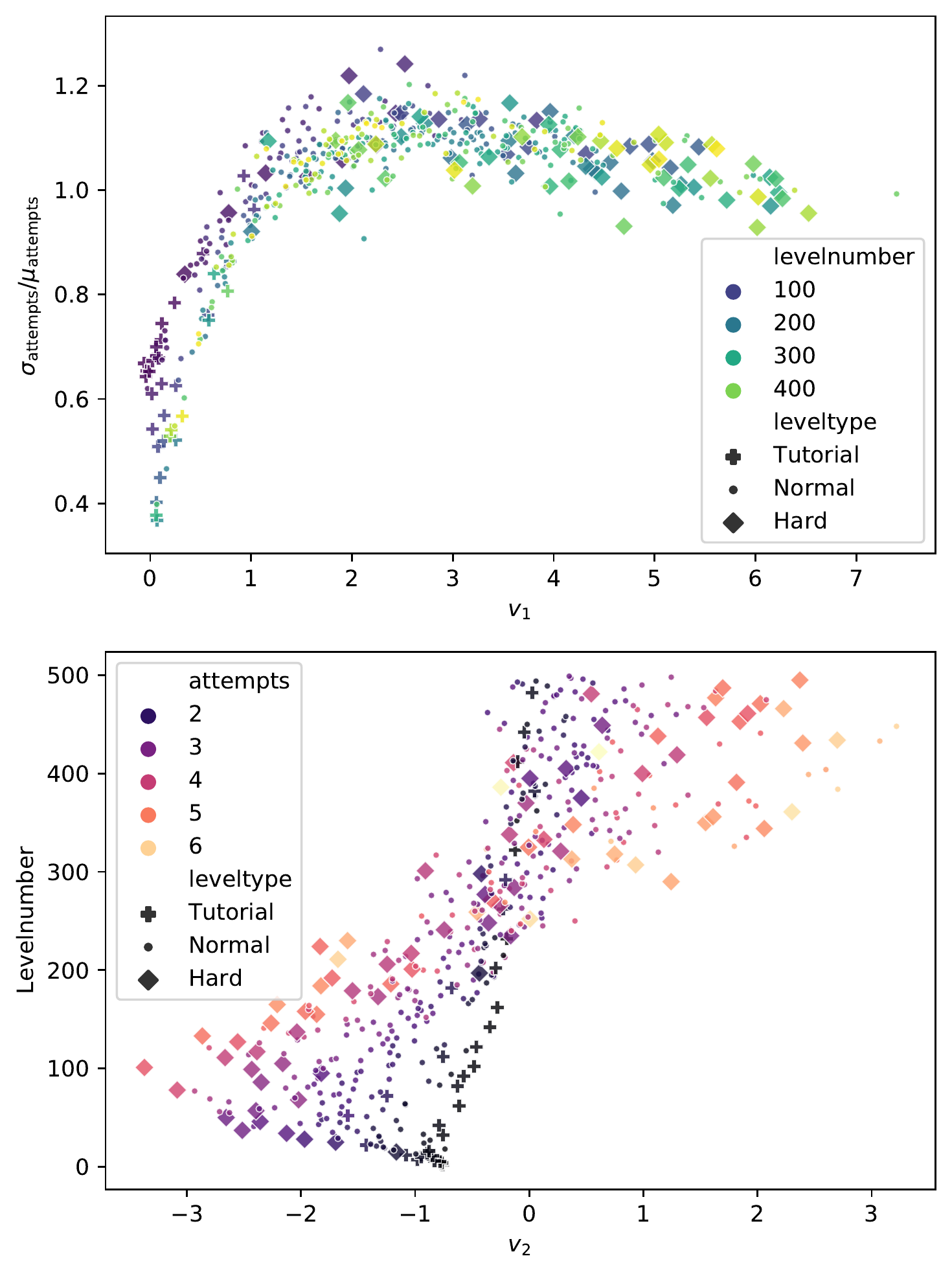}
    \caption{Top: Normalized variance versus the latent factors, $v_1$. Bottom: The level number plotted against the second latent factor, $v_2$. Note that level designers assign the \textit{hard} label based on their own initial estimates, so it does not always reflect whether a level requires many attempts on average.}
    \label{fig:variance vs v0}
    \Description[Scatterplot of other model parameters.]{Figure 10
Two scatterplots showing the latent factors, v1 and v2, for all levels plotted against the level variance in the top plot and level number in the bottom plot. 

Figure 10.top
Scatterplot of the first latent factor, v1. The x-axis represents values of the first latent factor, v1, and the y-axis the normalised variance in attempts per complete. The marker color expresses the level bands that each dot represents. The marker style indicates whether it is a normal, tutorial or hard level. The normalised variance starts at 0.3 and increases with increasing v1 until v1 equals 2. The normalised variance decreases slowly from 1.2 at v1=2 to 1.0 at v1=7. Most of the tutorial levels are concentrated in the lower left corner. The hard levels are more concentrated at higher v1 but appear throughout the whole v1 range. Hard levels all appear to have a high normalised variance. Overall, the plot shows that the first latent factor for the levels is related to the variance in attempts on the levels.

Figure 10.bottom
Scatterplot showing the second latent factor, v2, for the levels depending on the level number. The x-axis represents values of the second latent factor, v2, and the y-axis the level number in ascending order. The marker colors represent the average number of attempts per complete on the level. The marker style indicates whether it is a normal, tutorial or hard level. The tutorial levels are concentrated between v2 values from -1 to 0. The hard levels span the whole range of v2 from -3.5 to 3. From level 1 to around 250, v2 is negative, while v2 is positive after level 250. The more average attempts on a level, the larger its v2 is in magnitude.}
\end{figure}

\begin{figure}[t]
    \centering
    \begin{subfigure}{1\columnwidth}
    \includegraphics[trim={0.15cm 0 0.15cm 0},clip,width=0.49\columnwidth]{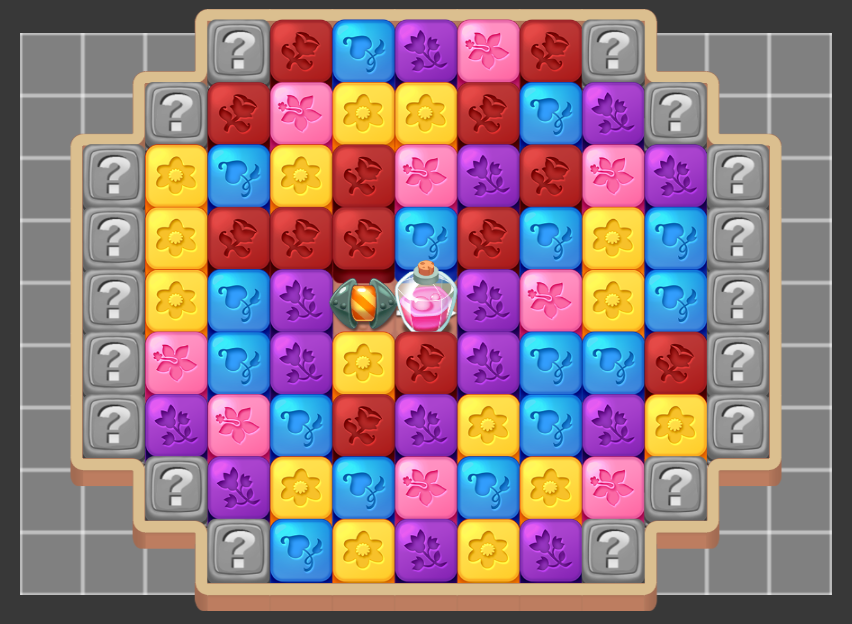}\hfill
    \includegraphics[width=0.49\columnwidth]{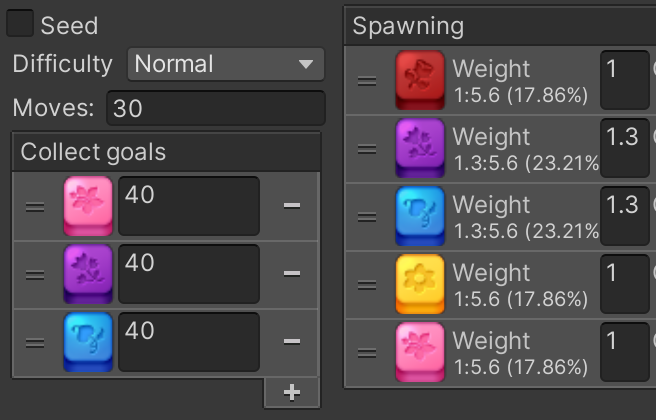}
    \caption{Level 5. The level is a tutorial level and has the 7th lowest $v_1$. It does not require any special strategies to complete.}
    \end{subfigure}
    \\[\smallskipamount]
    \begin{subfigure}{1\columnwidth}
    \includegraphics[width=0.49\columnwidth]{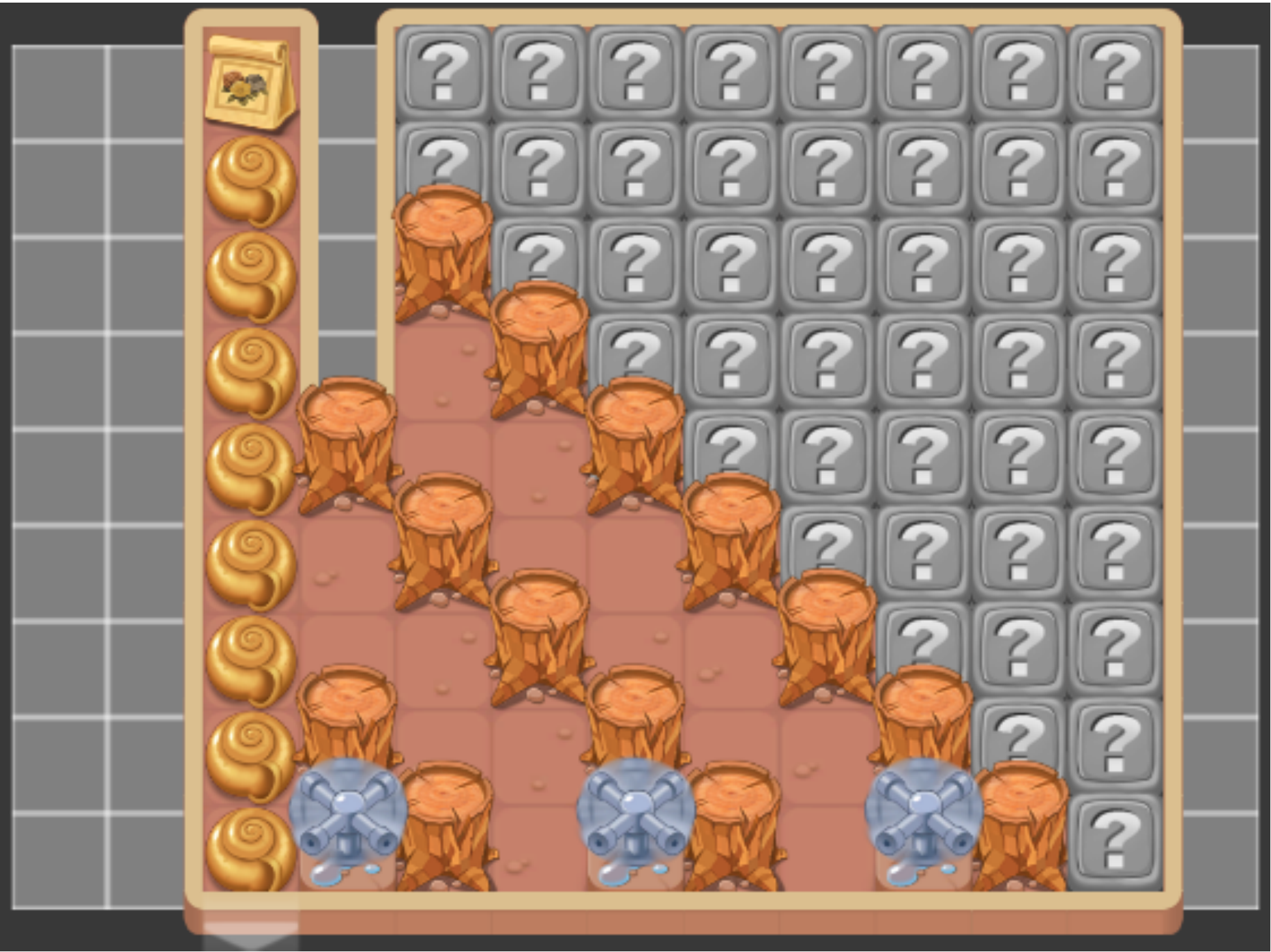}\hfill
    \includegraphics[width=0.49\columnwidth]{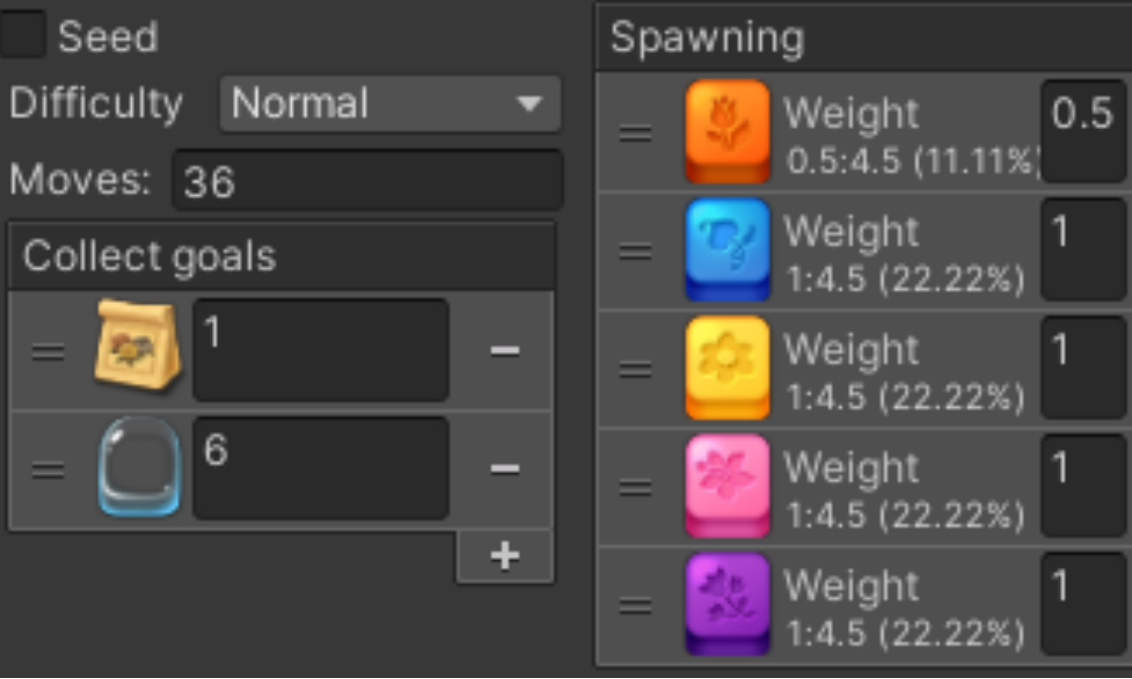}
    \caption{Level 383. The level is hard and has the highest $v_1$. In order to complete the level, the first level goal requires removing all the shells to the left, so the seed bag drops down to the bottom, while the second goal requires attacking either of the three bubble spawners at the bottom and then destroying six bubbles. The shells and tree stumps are both hard blockers, requiring power pieces for their removal.}
    \end{subfigure}
    \caption{Examples of a tutorial (\textit{a}) and hard level (\textit{b}). The left-hand side shows the level layout, and the right-hand side shows the move limit, level objectives, and color distributions.
    The pieces marked with '?' are assigned a random color at the start given by the weight distribution shown to the right.}
    \label{fig:level383 highest v0}
    \Description[Examples of an easy and hard level and their layout.]{Figure 11
Examples of a tutorial and a hard puzzle level in Lily's Garden. Each of the two figures consists of two subfigures, showing the level layout and goals.

Figure 11.a
The layout and goals of tutorial level 5 in Lily's Garden. The layout appears circular with many basic pieces having pre-assigned colors at the level start. In the middle is a powerful rocket+flask power piece combination. The player has to collect 40 pink, 40 purple and 40 blue basic pieces within 30 moves. There are 5 different colors that can appear (red, purple, blue, yellow, pink), with purple and blue pieces appearing slightly more frequently.

Figure 11.b
The layout and goals of level 383 in Lily's Garden, which can be considered hard. The layout is square and has no pre-assigned colors on the basic pieces at level start. There are tree stumps that block the lower left part of the level. The player has to collect 1 pink and 6 bubbles within 36 moves. There are 5 different colors that can appear (orange, blue, yellow, pink, purple), with orange pieces appearing less frequently.}
\end{figure}

To support this interpretation, we investigated levels at the extreme values of $v_1$. Since $v_1$ is strongly correlated with the average number of attempts to complete a level, which in turn is strongly correlated with the attempt variation on said level, we show the  attempt variation normalized by the average attempts as a function of $v_1$ in Fig. \ref{fig:variance vs v0}. At small values of $v_1$, we find the tutorial levels which generally have a high move limit and do not require any special strategies. Conversely, at large $v_1$ are the levels with the greatest difficulty and variance compared to the mean. As an illustrative example, we show level 383 in Fig.~\ref{fig:level383 highest v0}, which has the highest $v_1$ among all levels. There are multiple gameplay reasons why this level might have high variance and difficulty:
\begin{enumerate}
    \item The level contains blockers, which can only be destroyed with power pieces. Less skilled players will have a harder time creating the necessary pieces.
    \item The spawning weights on piece colors are almost uniform, making it harder to create power pieces.
    \item There is one specific power piece combination (2 magics, requiring 2 combinations of 10 pieces) that can easily win a level. More skilled players may have learned this strategy.
    \item The board itself is larger than average, allowing more possible strategies (e.g., focus on the top/bottom/left/right side).
\end{enumerate}

All these aspects lead to the perceived level difficulty being highly variable and dependable on the player's skill and game knowledge, supporting the interpretation that $v_1$ expresses how large an impact the player's skill can have on a given level. In contrast, level 5 in Fig.~\ref{fig:level383 highest v0} has the 7th lowest $v_1$ and requires a much more simple strategy to win, affording little variation in individual player difficulty.

\textbf{Second latent factor, $\mathbf{v_2}$:} 
This factor is strongly negatively correlated with the level number, as shown in Fig. \ref{fig:variance vs v0}. For levels below 250, $v_2$ tends to be negative and becomes positive beyond this threshold. There is nothing in the gameplay that changes drastically in the first 500 levels, but since the level number is linked to how many players have played the level, we interpret $v_2$ to capture aspects that are more temporal and related to the underlying data distribution: there are more observations on early levels, and different players have played later levels. It is thus likely that FM models with more factors find such spurious connections in the data.

For a deeper interpretation of $v_2$, we note that that $v_2$ is inversely correlated with player features as compared to $v_1$. 
Since $v_1$ can be interpreted as player consistency and skill, $v_2$ appears to capture similar aspects but with the opposite sign, \rebut{although it should be noted that the sign of the factors is arbitrary, as the dot product between user and content factors stays unchanged if one changes the sign of both factors.}
Since $v_2$ for the levels is centered around 0, the effect on the prediction depends on the player and their progress: a negative $v_2$ for a player means they spend more attempts on early levels but fewer attempts later on, while a positive $v_2$ signals a reduction in predicted attempts early on and more attempts later on. The $v_2$ factor can therefore also be related to a temporal shift in a player's playstyle compared to their peers. Crucially, this interpretation is less clear than for the first latent factor as $v_2$ appears to capture aspects more connected to the data collection rather than the player-level relation.

\section{Discussion}

Overall, our results from this case study suggest that FM outperforms both the naive baseline and RF, especially if augmented with similar features as RF.
Although RF performs better with fewer observations than non-augmented FM, the latter is game-agnostic, can be informative about the inter-dependency between players and levels, and does not require expert knowledge on relevant player and level descriptors. If such information is available, it can be utilized by FMs in addition to the player-level data.
Crucially though, there remain some caveats and limitations regarding the practical use of FMs and their generalizability, which we further elaborate on below.


\subsection{\rebut{New Players and Content}}

One issue with FMs is the cold start problem where the model does not learn model parameters ($w$, $\mathbf{v}$) for a specific player if they were not included in the training data (e.g.~if a model is trained during the night and the player starts playing the next day).
This can be mitigated by including additional player information, as we tested in this study.
While this increases computational complexity, the capacity to describe the player with a set of features that can be calculated immediately enables predictions without retraining the whole model.

While we did not extend our research to include unseen levels, the same cold start problem also exists in that case. We expect this to be more difficult to mitigate since information about the historical average number of attempts on a level -- a very strong predictor for the individualized predictions -- is unavailable, and the entropy of the color distribution (which \emph{is} available) only explains a small percentage of the variation. More work is therefore necessary to identify relevant level features, but some immediate options could be to include AI playtest agent data since this data can be strongly correlated with the player pass rate \cite{jeppe2020estimatinglearning,roohi2020predicting,roohi_predicting_2021}. We advocate FMs as a straightforward extension to the current AI playtesting approaches that will be able to capture the temporal and cohort differences, unlike previous prediction models.

\subsection{\rebut{Utility to Game Designers}}

A lot of work on DDA focuses on immediate adjustments to the game during play. 
This may be possible using an FM model and relational data; however, the strength of the FM approach is not just accurate predictions but also the modeling of further player characteristics. 
The latent factors were identified to be related to the players' skill and consistency over time, and this information can be relevant to many other features of the game. For instance, it could be used to cluster people together for in-game tournaments or groups so that they match in skill level and the competition is fair. It can also help level designers in proactively maintaining the level database by identifying problematic levels and bottlenecks before a new player cohort reaches this point. Ideally, in-game purchase data can also be incorporated into the process to provide estimates for both expected difficulty and monetization effectiveness and let level designers make informed decisions on necessary changes.

Before a tool like this can be implemented in a live game, though, there are a number of practical challenges that need to be considered first. Level design is very often an iterative process where minor adjustments are repeatedly performed on the level. Any change to the level will lead to a different difficulty, and the challenge is to convey this to the model. 
While some changes are easy to quantify, such as an alteration of the move limit or goals, others can be more tricky: even minor changes in the layout (e.g., assigning specific colors to start pieces or creating a hole in the level) can have a large impact on the difficulty, creating almost an entirely new level.
We argue that FMs should be able to deal with both cases: if the change is easily quantifiable, it can be added as relational data, and if not, the level can be counted as a completely new level.
While this would discard the information that two levels are very similar and increase data sparsity, FMs generally perform very well on sparse data compared to other algorithms, making them a very good candidate to use in the game industry.

\subsection{\rebut{Sensitivity to Randomness}}
As noted earlier, due to the stochastic nature of Lily's Garden, it is not possible to predict the exact number of attempts of a player on a level. An alternative could be to predict the probability of success per attempt, similar to Xue et al. \cite{xue_dynamic_2017}. Approaching the optimization problem from this perspective would require replacing the RMSE and MAE for evaluating the models with a measure such as the Poisson deviance, which is more appropriate for strictly positive counting data. However, for this study, we focus on RMSE and MAE, as these measures are more canonical and easier to interpret by game designers.


\subsection{\rebut{Generalizability to Other Games and Domains}}
\rebut{Our study is based on one specific puzzle game, but we expect our approach to work for other games as well, including other types of puzzle games, different genres (e.g.~platformers or first-person shooters), or even different types of games (e.g.~educational games).
FMs are task-agnostic and can be applied to any game where the following requirements are met:

\begin{enumerate}
    \item The variable of interest, e.g., pass rate or time taken, can be predicted as a multiplicative (or divisive) combination of user and content latent variables such as skill and difficulty.
    \item Multiple users are exposed to the same units of content such as puzzles or levels, allowing for inference of the latent variables.
\end{enumerate}

Many games meet these requirements, but there are notable exceptions, such as highly/completely random games, where the player has little effect on the outcome, and procedurally generated games, where the generated content can be unique for each player. 
Some instances of procedural content generation might still allow modeling latent player-content relations, e.g.~different players struggling with different enemies. 
However, this needs to be evaluated on a per-game basis. 


While we expect the model parameters to capture similar aspects across games, we do not consider it feasible to directly transfer a trained model to another game unless the games are very similar in terms of both gameplay and the distribution of predictor variables.
However, if the same players play multiple games, all the multi-game user-content interactions could, in principle, be combined into one big FM dataset. 
This is a potential direction for future research.
It should be noted, though, that games can have different types of challenges, such as emotional, cognitive, and physical challenges \cite{denisova_challenge_2017}.
To model all the challenge and skill facets of multiple games, one will likely need more FM factors than our default 2.}

\subsection{\rebut{Generalizability to Difficulty and/or Skill Varying Over Time}}
\rebut{Our approach does not explicitly take into account players' learning and skill improvement during play. 
However, Fig.~\ref{fig:baseline improvement nextlevels} demonstrates that the model performs well more than 100 levels into the future. Predicting beyond such a safe horizon is typically not necessary, as the model can be retrained periodically to include new observations of each player progressing through the game and gaining in skill. For instance, in our case, it would be feasible to retrain daily, as most players consume far less than 100 levels per day (median = 7, 90th quantile = 31), and the training takes only 14 hours on an Intel Cascade Lake \textit{c2-standard-16} Google Cloud machine.}

\rebut{Generalizability to dynamically varying skill and difficulty could also be improved with additional FM input features. An obvious choice is to utilize difficulty features such as a level's move limit or the amounts and types of enemies. It should also be possible to utilize player statistics such as total playtime, to allow FM to explicitly model skill acquisition over time. Future work is needed to investigate whether and how much such extra features improve prediction accuracy. } 

\section{Conclusion}
We have considered the problem of predicting personalized perceived difficulty and specifically how many attempts a player will spend on completing a level in a commercial mobile puzzle game. To this end, we compared four approaches: a simple non-personalized baseline, a Random Forest regressor, and Factorization Machines (FMs), the latter with and without game-specific features.

In terms of prediction performance, both the RF and FM methods achieved a lower MAE than the baseline after 10 and 30 observations of the players, respectively, answering our second research question about how many observations are necessary for being able to discern between players. The FM method that utilizes the same features as the RF model achieves better performance than all the other models already after 20 observations. All models had a lower RMSE than the baseline after 10 observations, indicating that any kind of personalized prediction can improve difficulty estimates.

A deeper analysis allowed us to answer our last research question about the FM parameters and their correlation with player and level characteristics. The first-order term, $w$, captures level difficulty, while the first latent factor, $v_1$, captures a player-level interaction that quantifies the player skill and level randomness: 
for levels, $v_1$ was strictly positive (except for 5 tutorial levels) and can be thought of as a level variance, whereas for players, both the magnitude and sign of $v_1$ indicate how sensitive and consistent a player is to the stochastic elements of a level and can thus be thought of as an expression of skill. Lastly, $v_2$ captures signals related to the underlying data distribution and temporal changes in player behavior.
For levels, $v_2$ is strongly correlated with the level number. The latent factor of the player, therefore, describes whether they perform better or worse over the progress of levels compared to their peers and, in a sense, how well they learn how to play. Including more latent factors led to overfitting and was thus not analyzed in more detail.

Overall, we find that FMs have multiple advantages over other prediction approaches: they outperform other typical approaches even when only relying on game-agnostic player-level-attempt data but have the option to utilize more fine-grained data to further improve performance; the FM model parameters provide interpretable results; FMs are scalable to large amounts of data.
FMs therefore show a large potential for difficulty modeling not just for research but also in an industrial context where such advantages are requirements. 

\section{Future Work}

We have focused on estimating how many attempts a player will spend on completing a level, but there are a number of extensions to the model that may make it even more useful for game designers.


The model performance clearly improves with additional information, especially when only a few player observations are available. A relevant line of future investigation is therefore to identify the features that have a large impact on the model performance and optimize the feature selection to prevent overfitting.
Additionally, including AI playtest agent data may also improve the predictions on levels with very few or no player-level observations.

The same player-level data could also be used in future work to predict variables other than the number of attempts. Other useful prediction targets could include churn, the probability of purchasing a continue, or the number of actions required to complete a level.
\rebut{Furthermore, the learned model parameters might be useful for other modeling approaches such as personalized offers, churn prediction, or sampling criteria for A/B testing new content to ensure enough player variation of the content.}

\begin{acks}
Many thanks to our anonymous reviewers for their extremely valuable feedback. CG has been partly funded by the Academy of Finland Flagship program \emph{Finnish Center for Artificial Intelligence} (FCAI). We thank Tactile Games for supporting this work, and especially the level design team for sharing their expertise and insights.
\end{acks}

\bibliographystyle{ACM-Reference-Format}
\bibliography{references}


\begin{thebibliography}{56}


\ifx \showCODEN    \undefined \def \showCODEN     #1{\unskip}     \fi
\ifx \showDOI      \undefined \def \showDOI       #1{#1}\fi
\ifx \showISBNx    \undefined \def \showISBNx     #1{\unskip}     \fi
\ifx \showISBNxiii \undefined \def \showISBNxiii  #1{\unskip}     \fi
\ifx \showISSN     \undefined \def \showISSN      #1{\unskip}     \fi
\ifx \showLCCN     \undefined \def \showLCCN      #1{\unskip}     \fi
\ifx \shownote     \undefined \def \shownote      #1{#1}          \fi
\ifx \showarticletitle \undefined \def \showarticletitle #1{#1}   \fi
\ifx \showURL      \undefined \def \showURL       {\relax}        \fi
\providecommand\bibfield[2]{#2}
\providecommand\bibinfo[2]{#2}
\providecommand\natexlab[1]{#1}
\providecommand\showeprint[2][]{arXiv:#2}

\bibitem[Alexander et~al\mbox{.}(2013)]%
        {alexander_investigation_2013}
\bibfield{author}{\bibinfo{person}{Justin~T. Alexander}, \bibinfo{person}{John
  Sear}, {and} \bibinfo{person}{Andreas Oikonomou}.}
  \bibinfo{year}{2013}\natexlab{}.
\newblock \showarticletitle{An investigation of the effects of game difficulty
  on player enjoyment}.
\newblock \bibinfo{journal}{\emph{Entertainment Computing}}
  \bibinfo{volume}{4}, \bibinfo{number}{1} (\bibinfo{date}{Feb.}
  \bibinfo{year}{2013}), \bibinfo{pages}{53--62}.
\newblock
\showISSN{18759521}
\urldef\tempurl%
\url{https://doi.org/10.1016/j.entcom.2012.09.001}
\showDOI{\tempurl}


\bibitem[Anderson et~al\mbox{.}(2017)]%
        {anderson2017assessing}
\bibfield{author}{\bibinfo{person}{Ashton Anderson}, \bibinfo{person}{Jon
  Kleinberg}, {and} \bibinfo{person}{Sendhil Mullainathan}.}
  \bibinfo{year}{2017}\natexlab{}.
\newblock \showarticletitle{Assessing human error against a benchmark of
  perfection}.
\newblock \bibinfo{journal}{\emph{ACM Transactions on Knowledge Discovery from
  Data (TKDD)}} \bibinfo{volume}{11}, \bibinfo{number}{4}
  (\bibinfo{year}{2017}), \bibinfo{pages}{1--25}.
\newblock


\bibitem[Anwar et~al\mbox{.}(2017)]%
        {anwar_game_2017}
\bibfield{author}{\bibinfo{person}{Syed~Muhammad Anwar}, \bibinfo{person}{Talha
  Shahzad}, \bibinfo{person}{Zunaira Sattar}, \bibinfo{person}{Rahma Khan},
  {and} \bibinfo{person}{Muhammad Majid}.} \bibinfo{year}{2017}\natexlab{}.
\newblock \showarticletitle{A game recommender system using collaborative
  filtering (GAMBIT)}. In \bibinfo{booktitle}{\emph{2017 14th International
  Bhurban Conference on Applied Sciences and Technology (IBCAST)}}.
  \bibinfo{pages}{328--332}.
\newblock
\urldef\tempurl%
\url{https://doi.org/10.1109/IBCAST.2017.7868073}
\showDOI{\tempurl}


\bibitem[Brockmyer et~al\mbox{.}(2009)]%
        {brockmyer2009development}
\bibfield{author}{\bibinfo{person}{Jeanne~H Brockmyer},
  \bibinfo{person}{Christine~M Fox}, \bibinfo{person}{Kathleen~A Curtiss},
  \bibinfo{person}{Evan McBroom}, \bibinfo{person}{Kimberly~M Burkhart}, {and}
  \bibinfo{person}{Jacquelyn~N Pidruzny}.} \bibinfo{year}{2009}\natexlab{}.
\newblock \showarticletitle{{The Development of the Game Engagement
  Questionnaire: A Measure of Engagement in Video Game-Playing}}.
\newblock \bibinfo{journal}{\emph{Journal of Experimental Social Psychology}}
  \bibinfo{volume}{45}, \bibinfo{number}{4} (\bibinfo{year}{2009}),
  \bibinfo{pages}{624--634}.
\newblock


\bibitem[Bunian et~al\mbox{.}(2017)]%
        {bunian_modeling_nodate}
\bibfield{author}{\bibinfo{person}{Sara Bunian}, \bibinfo{person}{Alessandro
  Canossa}, \bibinfo{person}{Randy Colvin}, {and} \bibinfo{person}{Magy~Seif
  El-Nasr}.} \bibinfo{year}{2017}\natexlab{}.
\newblock \showarticletitle{Modeling individual differences in game behavior
  using HMM}.
\newblock  (\bibinfo{year}{2017}).
\newblock


\bibitem[Cairns(2016)]%
        {cairns2016engagement}
\bibfield{author}{\bibinfo{person}{Paul Cairns}.}
  \bibinfo{year}{2016}\natexlab{}.
\newblock \showarticletitle{Engagement in Digital Games}.
\newblock In \bibinfo{booktitle}{\emph{Why Engagement Matters:
  Cross-Disciplinary Perspectives of User Engagement in Digital Media}},
  \bibfield{editor}{\bibinfo{person}{Heather O'Brien} {and}
  \bibinfo{person}{Paul Cairns}} (Eds.). \bibinfo{publisher}{Springer
  International Publishing}, \bibinfo{address}{Cham}, \bibinfo{pages}{81--104}.
\newblock
\showISBNx{978-3-319-27446-1}
\urldef\tempurl%
\url{https://doi.org/10.1007/978-3-319-27446-1_4}
\showDOI{\tempurl}


\bibitem[Cen et~al\mbox{.}(2006)]%
        {cen2006learning}
\bibfield{author}{\bibinfo{person}{Hao Cen}, \bibinfo{person}{Kenneth
  Koedinger}, {and} \bibinfo{person}{Brian Junker}.}
  \bibinfo{year}{2006}\natexlab{}.
\newblock \showarticletitle{Learning factors analysis--a general method for
  cognitive model evaluation and improvement}. In
  \bibinfo{booktitle}{\emph{International conference on intelligent tutoring
  systems}}. Springer, \bibinfo{pages}{164--175}.
\newblock


\bibitem[Chen(2007)]%
        {chen_flow_2007}
\bibfield{author}{\bibinfo{person}{Jenova Chen}.}
  \bibinfo{year}{2007}\natexlab{}.
\newblock \showarticletitle{Flow in Games (and Everything Else)}.
\newblock \bibinfo{journal}{\emph{Commun. ACM}} \bibinfo{volume}{50},
  \bibinfo{number}{4} (\bibinfo{date}{apr} \bibinfo{year}{2007}),
  \bibinfo{pages}{31–34}.
\newblock
\showISSN{0001-0782}
\urldef\tempurl%
\url{https://doi.org/10.1145/1232743.1232769}
\showDOI{\tempurl}


\bibitem[Cheuque et~al\mbox{.}(2019)]%
        {cheuque_recommender_2019}
\bibfield{author}{\bibinfo{person}{Germán Cheuque}, \bibinfo{person}{José
  Guzmán}, {and} \bibinfo{person}{Denis Parra}.}
  \bibinfo{year}{2019}\natexlab{}.
\newblock \showarticletitle{Recommender {Systems} for {Online} {Video} {Game}
  {Platforms}: the {Case} of {STEAM}}. In \bibinfo{booktitle}{\emph{Companion
  {Proceedings} of {The} 2019 {World} {Wide} {Web} {Conference}}}.
  \bibinfo{publisher}{ACM}, \bibinfo{address}{San Francisco USA},
  \bibinfo{pages}{763--771}.
\newblock
\showISBNx{978-1-4503-6675-5}
\urldef\tempurl%
\url{https://doi.org/10.1145/3308560.3316457}
\showDOI{\tempurl}


\bibitem[Csikszentmihalyi and Csikzentmihaly(1990)]%
        {csikszentmihalyi1990flow}
\bibfield{author}{\bibinfo{person}{Mihaly Csikszentmihalyi} {and}
  \bibinfo{person}{Mihaly Csikzentmihaly}.} \bibinfo{year}{1990}\natexlab{}.
\newblock \bibinfo{booktitle}{\emph{Flow: The psychology of optimal
  experience}}. Vol.~\bibinfo{volume}{1990}.
\newblock \bibinfo{publisher}{Harper \& Row New York}.
\newblock


\bibitem[Denisova et~al\mbox{.}(2020)]%
        {denisova_measuring_2020}
\bibfield{author}{\bibinfo{person}{Alena Denisova}, \bibinfo{person}{Paul
  Cairns}, \bibinfo{person}{Christian Guckelsberger}, {and}
  \bibinfo{person}{David Zendle}.} \bibinfo{year}{2020}\natexlab{}.
\newblock \showarticletitle{Measuring perceived challenge in digital games:
  {Development} \& validation of the challenge originating from recent gameplay
  interaction scale ({CORGIS})}.
\newblock \bibinfo{journal}{\emph{International Journal of Human-Computer
  Studies}}  \bibinfo{volume}{137} (\bibinfo{date}{May} \bibinfo{year}{2020}),
  \bibinfo{pages}{102383}.
\newblock
\showISSN{10715819}
\urldef\tempurl%
\url{https://doi.org/10.1016/j.ijhcs.2019.102383}
\showDOI{\tempurl}


\bibitem[Denisova et~al\mbox{.}(2017)]%
        {denisova_challenge_2017}
\bibfield{author}{\bibinfo{person}{Alena Denisova}, \bibinfo{person}{Christian
  Guckelsberger}, {and} \bibinfo{person}{David Zendle}.}
  \bibinfo{year}{2017}\natexlab{}.
\newblock \showarticletitle{Challenge in {Digital} {Games}: {Towards}
  {Developing} a {Measurement} {Tool}}. In
  \bibinfo{booktitle}{\emph{Proceedings of the 2017 {CHI} {Conference}
  {Extended} {Abstracts} on {Human} {Factors} in {Computing} {Systems}}}.
  \bibinfo{publisher}{ACM}, \bibinfo{address}{Denver Colorado USA},
  \bibinfo{pages}{2511--2519}.
\newblock
\showISBNx{978-1-4503-4656-6}
\urldef\tempurl%
\url{https://doi.org/10.1145/3027063.3053209}
\showDOI{\tempurl}


\bibitem[Drachen et~al\mbox{.}(2016)]%
        {drachen_rapid_nodate}
\bibfield{author}{\bibinfo{person}{Anders Drachen},
  \bibinfo{person}{Eric~Thurston Lundquist}, \bibinfo{person}{Yungjen Kung},
  \bibinfo{person}{Pranav Rao}, \bibinfo{person}{Rafet Sifa},
  \bibinfo{person}{Julian Runge}, {and} \bibinfo{person}{Diego Klabjan}.}
  \bibinfo{year}{2016}\natexlab{}.
\newblock \showarticletitle{Rapid prediction of player retention in
  free-to-play mobile games}.
\newblock  (\bibinfo{year}{2016}).
\newblock


\bibitem[Drey et~al\mbox{.}(2021)]%
        {drey_be_2021}
\bibfield{author}{\bibinfo{person}{Tobias Drey}, \bibinfo{person}{Fabian
  Fischbach}, \bibinfo{person}{Pascal Jansen}, \bibinfo{person}{Julian
  Frommel}, \bibinfo{person}{Michael Rietzler}, {and} \bibinfo{person}{Enrico
  Rukzio}.} \bibinfo{year}{2021}\natexlab{}.
\newblock \showarticletitle{To {Be} or {Not} to {Be} {Stuck}, or {Is} {It} a
  {Continuum}?: {A} {Systematic} {Literature} {Review} on the {Concept} of
  {Being} {Stuck} in {Games}}.
\newblock \bibinfo{journal}{\emph{Proceedings of the ACM on Human-Computer
  Interaction}} \bibinfo{volume}{5}, \bibinfo{number}{CHI PLAY}
  (\bibinfo{date}{Oct.} \bibinfo{year}{2021}), \bibinfo{pages}{1--35}.
\newblock
\showISSN{2573-0142}
\urldef\tempurl%
\url{https://doi.org/10.1145/3474656}
\showDOI{\tempurl}


\bibitem[Goldin et~al\mbox{.}(2018)]%
        {goldin2018most}
\bibfield{author}{\bibinfo{person}{Ilya Goldin}, \bibinfo{person}{April
  Galyardt}, {et~al\mbox{.}}} \bibinfo{year}{2018}\natexlab{}.
\newblock \showarticletitle{Most of the Time, It Works Every Time: Limitations
  in Refining Domain Models with Learning Curves}.
\newblock \bibinfo{journal}{\emph{Journal of Educational Data Mining}}
  \bibinfo{volume}{10}, \bibinfo{number}{2} (\bibinfo{year}{2018}),
  \bibinfo{pages}{55--92}.
\newblock


\bibitem[Gonzalez-Duque et~al\mbox{.}(2021)]%
        {gonzalezduque2021fastmodelling}
\bibfield{author}{\bibinfo{person}{Miguel Gonzalez-Duque},
  \bibinfo{person}{Rasmus~Berg Palm}, {and} \bibinfo{person}{Sebastian Risi}.}
  \bibinfo{year}{2021}\natexlab{}.
\newblock \showarticletitle{Fast Game Content Adaptation Through Bayesian-based
  Player Modelling}. In \bibinfo{booktitle}{\emph{2021 IEEE Conference on Games
  (CoG)}}. \bibinfo{pages}{01--08}.
\newblock
\urldef\tempurl%
\url{https://doi.org/10.1109/CoG52621.2021.9619018}
\showDOI{\tempurl}


\bibitem[González-Duque et~al\mbox{.}(2020)]%
        {gonzalezduque2020findingerror}
\bibfield{author}{\bibinfo{person}{Miguel González-Duque},
  \bibinfo{person}{Rasmus~Berg Palm}, \bibinfo{person}{David Ha}, {and}
  \bibinfo{person}{Sebastian Risi}.} \bibinfo{year}{2020}\natexlab{}.
\newblock \showarticletitle{Finding Game Levels with the Right Difficulty in a
  Few Trials through Intelligent Trial-and-Error}. In
  \bibinfo{booktitle}{\emph{2020 IEEE Conference on Games (CoG)}}.
  \bibinfo{pages}{503--510}.
\newblock
\urldef\tempurl%
\url{https://doi.org/10.1109/CoG47356.2020.9231548}
\showDOI{\tempurl}


\bibitem[Gudmundsson et~al\mbox{.}(2018)]%
        {gudmundsson2018human}
\bibfield{author}{\bibinfo{person}{Stefan~Freyr Gudmundsson},
  \bibinfo{person}{Philipp Eisen}, \bibinfo{person}{Erik Poromaa},
  \bibinfo{person}{Alex Nodet}, \bibinfo{person}{Sami Purmonen},
  \bibinfo{person}{Bartlomiej Kozakowski}, \bibinfo{person}{Richard Meurling},
  {and} \bibinfo{person}{Lele Cao}.} \bibinfo{year}{2018}\natexlab{}.
\newblock \showarticletitle{Human-like playtesting with deep learning}. In
  \bibinfo{booktitle}{\emph{2018 IEEE Conference on Computational Intelligence
  and Games (CIG)}}. IEEE, \bibinfo{pages}{1--8}.
\newblock


\bibitem[Harpstead and Aleven(2015)]%
        {hardstead2015usinggame}
\bibfield{author}{\bibinfo{person}{Erik Harpstead} {and}
  \bibinfo{person}{Vincent Aleven}.} \bibinfo{year}{2015}\natexlab{}.
\newblock \showarticletitle{Using Empirical Learning Curve Analysis to Inform
  Design in an Educational Game}. In \bibinfo{booktitle}{\emph{Proceedings of
  the 2015 Annual Symposium on Computer-Human Interaction in Play}} (London,
  United Kingdom) \emph{(\bibinfo{series}{CHI PLAY '15})}.
  \bibinfo{publisher}{Association for Computing Machinery},
  \bibinfo{address}{New York, NY, USA}, \bibinfo{pages}{197–207}.
\newblock
\showISBNx{9781450334662}
\urldef\tempurl%
\url{https://doi.org/10.1145/2793107.2793128}
\showDOI{\tempurl}


\bibitem[Hong et~al\mbox{.}(2019)]%
        {hong2019interaction}
\bibfield{author}{\bibinfo{person}{Fuxing Hong}, \bibinfo{person}{Dongbo
  Huang}, {and} \bibinfo{person}{Ge Chen}.} \bibinfo{year}{2019}\natexlab{}.
\newblock \showarticletitle{Interaction-aware factorization machines for
  recommender systems}. In \bibinfo{booktitle}{\emph{Proceedings of the AAAI
  Conference on Artificial Intelligence}}, Vol.~\bibinfo{volume}{33}.
  \bibinfo{pages}{3804--3811}.
\newblock


\bibitem[Jennings-Teats et~al\mbox{.}(2010)]%
        {jennings2010polymorph}
\bibfield{author}{\bibinfo{person}{Martin Jennings-Teats},
  \bibinfo{person}{Gillian Smith}, {and} \bibinfo{person}{Noah Wardrip-Fruin}.}
  \bibinfo{year}{2010}\natexlab{}.
\newblock \showarticletitle{Polymorph: dynamic difficulty adjustment through
  level generation}. In \bibinfo{booktitle}{\emph{Proceedings of the 2010
  Workshop on Procedural Content Generation in Games}}. \bibinfo{pages}{1--4}.
\newblock


\bibitem[Kamaldinov and Makarov(2019)]%
        {kamaldinov2019deep}
\bibfield{author}{\bibinfo{person}{Ildar Kamaldinov} {and}
  \bibinfo{person}{Ilya Makarov}.} \bibinfo{year}{2019}\natexlab{}.
\newblock \showarticletitle{Deep reinforcement learning in match-3 game}. In
  \bibinfo{booktitle}{\emph{2019 IEEE conference on games (CoG)}}. IEEE,
  \bibinfo{pages}{1--4}.
\newblock


\bibitem[Kristensen and Burelli(2019)]%
        {kristensen_combining_2019}
\bibfield{author}{\bibinfo{person}{Jeppe~Theiss Kristensen} {and}
  \bibinfo{person}{Paolo Burelli}.} \bibinfo{year}{2019}\natexlab{}.
\newblock \showarticletitle{Combining {Sequential} and {Aggregated} {Data} for
  {Churn} {Prediction} in {Casual} {Freemium} {Games}}. In
  \bibinfo{booktitle}{\emph{2019 {IEEE} {Conference} on {Games} ({CoG})}}.
  \bibinfo{pages}{1--8}.
\newblock
\urldef\tempurl%
\url{https://doi.org/10.1109/CIG.2019.8848106}
\showDOI{\tempurl}
\newblock
\shownote{ISSN: 2325-4289}.


\bibitem[Kristensen et~al\mbox{.}(2020)]%
        {jeppe2020estimatinglearning}
\bibfield{author}{\bibinfo{person}{Jeppe~Theiss Kristensen},
  \bibinfo{person}{Arturo Valdivia}, {and} \bibinfo{person}{Paolo Burelli}.}
  \bibinfo{year}{2020}\natexlab{}.
\newblock \showarticletitle{Estimating Player Completion Rate in Mobile Puzzle
  Games Using Reinforcement Learning}. In \bibinfo{booktitle}{\emph{2020 IEEE
  Conference on Games (CoG)}}. \bibinfo{pages}{636--639}.
\newblock
\urldef\tempurl%
\url{https://doi.org/10.1109/CoG47356.2020.9231581}
\showDOI{\tempurl}


\bibitem[Kristensen et~al\mbox{.}(2021)]%
        {kristensen2021statistical}
\bibfield{author}{\bibinfo{person}{Jeppe~Theiss Kristensen},
  \bibinfo{person}{Arturo Valdivia}, {and} \bibinfo{person}{Paolo Burelli}.}
  \bibinfo{year}{2021}\natexlab{}.
\newblock \showarticletitle{Statistical Modelling of Level Difficulty in Puzzle
  Games}. In \bibinfo{booktitle}{\emph{2021 IEEE Conference on Games (CoG)}}.
  IEEE, \bibinfo{pages}{1--8}.
\newblock


\bibitem[Lazzaro(2004)]%
        {lazzaro2004we}
\bibfield{author}{\bibinfo{person}{R Lazzaro}.}
  \bibinfo{year}{2004}\natexlab{}.
\newblock \showarticletitle{Why we play games: 4 keys to more emotion}.
\newblock \bibinfo{journal}{\emph{Proc. Game Developers Conference 2004}}.
\newblock
\urldef\tempurl%
\url{https://cir.nii.ac.jp/crid/1572543025651858816}
\showURL{%
\tempurl}


\bibitem[Li et~al\mbox{.}(2021)]%
        {li_difficulty-aware_2021}
\bibfield{author}{\bibinfo{person}{Jiayu Li}, \bibinfo{person}{Hongyu Lu},
  \bibinfo{person}{Chenyang Wang}, \bibinfo{person}{Weizhi Ma},
  \bibinfo{person}{Min Zhang}, \bibinfo{person}{Xiangyu Zhao},
  \bibinfo{person}{Wei Qi}, \bibinfo{person}{Yiqun Liu}, {and}
  \bibinfo{person}{Shaoping Ma}.} \bibinfo{year}{2021}\natexlab{}.
\newblock \showarticletitle{A {Difficulty}-{Aware} {Framework} for {Churn}
  {Prediction} and {Intervention} in {Games}}. In
  \bibinfo{booktitle}{\emph{Proceedings of the 27th {ACM} {SIGKDD} {Conference}
  on {Knowledge} {Discovery} \& {Data} {Mining}}}. \bibinfo{publisher}{ACM},
  \bibinfo{address}{Virtual Event Singapore}, \bibinfo{pages}{943--952}.
\newblock
\showISBNx{978-1-4503-8332-5}
\urldef\tempurl%
\url{https://doi.org/10.1145/3447548.3467277}
\showDOI{\tempurl}


\bibitem[Linehan et~al\mbox{.}(2014)]%
        {linehan_learning_2014}
\bibfield{author}{\bibinfo{person}{Conor Linehan}, \bibinfo{person}{George
  Bellord}, \bibinfo{person}{Ben Kirman}, \bibinfo{person}{Zachary~H. Morford},
  {and} \bibinfo{person}{Bryan Roche}.} \bibinfo{year}{2014}\natexlab{}.
\newblock \showarticletitle{Learning curves: analysing pace and challenge in
  four successful puzzle games}. In \bibinfo{booktitle}{\emph{Proceedings of
  the first {ACM} {SIGCHI} annual symposium on {Computer}-human interaction in
  play}}. \bibinfo{publisher}{ACM}, \bibinfo{address}{Toronto Ontario Canada},
  \bibinfo{pages}{181--190}.
\newblock
\showISBNx{978-1-4503-3014-5}
\urldef\tempurl%
\url{https://doi.org/10.1145/2658537.2658695}
\showDOI{\tempurl}


\bibitem[Lora et~al\mbox{.}(2016)]%
        {lora2016dynamic}
\bibfield{author}{\bibinfo{person}{Diana Lora}, \bibinfo{person}{Antonio~A
  S{\'a}nchez-Ruiz}, \bibinfo{person}{Pedro~A Gonz{\'a}lez-Calero}, {and}
  \bibinfo{person}{Marco~A G{\'o}mez-Mart{\'\i}n}.}
  \bibinfo{year}{2016}\natexlab{}.
\newblock \showarticletitle{Dynamic difficulty adjustment in tetris}. In
  \bibinfo{booktitle}{\emph{The Twenty-Ninth International Flairs Conference}}.
\newblock


\bibitem[Mahlmann et~al\mbox{.}(2010)]%
        {mahlmann_predicting_2010}
\bibfield{author}{\bibinfo{person}{Tobias Mahlmann}, \bibinfo{person}{Anders
  Drachen}, \bibinfo{person}{Julian Togelius}, \bibinfo{person}{Alessandro
  Canossa}, {and} \bibinfo{person}{Georgios~N. Yannakakis}.}
  \bibinfo{year}{2010}\natexlab{}.
\newblock \showarticletitle{Predicting player behavior in {Tomb} {Raider}:
  {Underworld}}. In \bibinfo{booktitle}{\emph{Proceedings of the 2010 {IEEE}
  {Conference} on {Computational} {Intelligence} and {Games}}}.
  \bibinfo{pages}{178--185}.
\newblock
\urldef\tempurl%
\url{https://doi.org/10.1109/ITW.2010.5593355}
\showDOI{\tempurl}
\newblock
\shownote{ISSN: 2325-4289}.


\bibitem[Moon and Seo(2020)]%
        {moon2020dynamic}
\bibfield{author}{\bibinfo{person}{Hee-Seung Moon} {and} \bibinfo{person}{Jiwon
  Seo}.} \bibinfo{year}{2020}\natexlab{}.
\newblock \showarticletitle{Dynamic difficulty adjustment via fast user
  adaptation}. In \bibinfo{booktitle}{\emph{Adjunct Publication of the 33rd
  Annual ACM Symposium on User Interface Software and Technology}}.
  \bibinfo{pages}{13--15}.
\newblock


\bibitem[Mourato et~al\mbox{.}(2014)]%
        {mourato2014difficulty}
\bibfield{author}{\bibinfo{person}{Fausto Mourato}, \bibinfo{person}{Fernando
  Birra}, {and} \bibinfo{person}{Manuel~Pr{\'o}spero dos Santos}.}
  \bibinfo{year}{2014}\natexlab{}.
\newblock \showarticletitle{Difficulty in action based challenges: success
  prediction, players' strategies and profiling}. In
  \bibinfo{booktitle}{\emph{Proceedings of the 11th Conference on Advances in
  Computer Entertainment Technology}}. \bibinfo{pages}{1--10}.
\newblock


\bibitem[Or et~al\mbox{.}(2021)]%
        {or2021dl}
\bibfield{author}{\bibinfo{person}{Dvir~Ben Or}, \bibinfo{person}{Michael
  Kolomenkin}, {and} \bibinfo{person}{Gil Shabat}.}
  \bibinfo{year}{2021}\natexlab{}.
\newblock \showarticletitle{DL-DDA-Deep Learning based Dynamic Difficulty
  Adjustment with UX and Gameplay constraints}. In
  \bibinfo{booktitle}{\emph{2021 IEEE Conference on Games (CoG)}}. IEEE,
  \bibinfo{pages}{1--7}.
\newblock


\bibitem[Pedregosa et~al\mbox{.}(2011)]%
        {scikit-learn}
\bibfield{author}{\bibinfo{person}{F. Pedregosa}, \bibinfo{person}{G.
  Varoquaux}, \bibinfo{person}{A. Gramfort}, \bibinfo{person}{V. Michel},
  \bibinfo{person}{B. Thirion}, \bibinfo{person}{O. Grisel},
  \bibinfo{person}{M. Blondel}, \bibinfo{person}{P. Prettenhofer},
  \bibinfo{person}{R. Weiss}, \bibinfo{person}{V. Dubourg}, \bibinfo{person}{J.
  Vanderplas}, \bibinfo{person}{A. Passos}, \bibinfo{person}{D. Cournapeau},
  \bibinfo{person}{M. Brucher}, \bibinfo{person}{M. Perrot}, {and}
  \bibinfo{person}{E. Duchesnay}.} \bibinfo{year}{2011}\natexlab{}.
\newblock \showarticletitle{Scikit-learn: Machine Learning in {P}ython}.
\newblock \bibinfo{journal}{\emph{Journal of Machine Learning Research}}
  \bibinfo{volume}{12} (\bibinfo{year}{2011}), \bibinfo{pages}{2825--2830}.
\newblock


\bibitem[Pfau et~al\mbox{.}(2020)]%
        {pfau2020enemy}
\bibfield{author}{\bibinfo{person}{Johannes Pfau}, \bibinfo{person}{Jan~David
  Smeddinck}, {and} \bibinfo{person}{Rainer Malaka}.}
  \bibinfo{year}{2020}\natexlab{}.
\newblock \showarticletitle{Enemy within: Long-term motivation effects of deep
  player behavior models for dynamic difficulty adjustment}. In
  \bibinfo{booktitle}{\emph{Proceedings of the 2020 CHI Conference on Human
  Factors in Computing Systems}}. \bibinfo{pages}{1--10}.
\newblock


\bibitem[Power et~al\mbox{.}(2019)]%
        {power2019lost}
\bibfield{author}{\bibinfo{person}{Christopher Power}, \bibinfo{person}{Paul
  Cairns}, \bibinfo{person}{Alena Denisova}, \bibinfo{person}{Themis
  Papaioannou}, {and} \bibinfo{person}{Ruth Gultom}.}
  \bibinfo{year}{2019}\natexlab{}.
\newblock \showarticletitle{Lost at the edge of uncertainty: Measuring player
  uncertainty in digital games}.
\newblock \bibinfo{journal}{\emph{International Journal of Human--Computer
  Interaction}} \bibinfo{volume}{35}, \bibinfo{number}{12}
  (\bibinfo{year}{2019}), \bibinfo{pages}{1033--1045}.
\newblock


\bibitem[Pusey et~al\mbox{.}(2021)]%
        {pusey_puzzle_2021}
\bibfield{author}{\bibinfo{person}{Megan Pusey}, \bibinfo{person}{Kok~Wai
  Wong}, {and} \bibinfo{person}{Natasha~Anne Rappa}.}
  \bibinfo{year}{2021}\natexlab{}.
\newblock \showarticletitle{The {Puzzle} {Challenge} {Analysis} {Tool}. {A}
  {Tool} for {Analysing} the {Cognitive} {Challenge} {Level} of {Puzzles} in
  {Video} {Games}}.
\newblock \bibinfo{journal}{\emph{Proceedings of the ACM on Human-Computer
  Interaction}} \bibinfo{volume}{5}, \bibinfo{number}{CHI PLAY}
  (\bibinfo{date}{Oct.} \bibinfo{year}{2021}), \bibinfo{pages}{1--27}.
\newblock
\showISSN{2573-0142}
\urldef\tempurl%
\url{https://doi.org/10.1145/3474703}
\showDOI{\tempurl}


\bibitem[Ren et~al\mbox{.}(2019)]%
        {ren_grade_2019}
\bibfield{author}{\bibinfo{person}{Zhiyun Ren}, \bibinfo{person}{Xia Ning},
  \bibinfo{person}{Andrew~S. Lan}, {and} \bibinfo{person}{Huzefa Rangwala}.}
  \bibinfo{year}{2019}\natexlab{}.
\newblock \showarticletitle{Grade {Prediction} with {Neural} {Collaborative}
  {Filtering}}. In \bibinfo{booktitle}{\emph{2019 {IEEE} {International}
  {Conference} on {Data} {Science} and {Advanced} {Analytics} ({DSAA})}}.
  \bibinfo{pages}{1--10}.
\newblock
\urldef\tempurl%
\url{https://doi.org/10.1109/DSAA.2019.00014}
\showDOI{\tempurl}


\bibitem[Rendle(2010)]%
        {rendle2010factorization}
\bibfield{author}{\bibinfo{person}{Steffen Rendle}.}
  \bibinfo{year}{2010}\natexlab{}.
\newblock \showarticletitle{Factorization machines}. In
  \bibinfo{booktitle}{\emph{2010 IEEE International conference on data
  mining}}. IEEE, \bibinfo{pages}{995--1000}.
\newblock


\bibitem[Rendle(2012)]%
        {rendle_factorization_2012}
\bibfield{author}{\bibinfo{person}{Steffen Rendle}.}
  \bibinfo{year}{2012}\natexlab{}.
\newblock \showarticletitle{Factorization {Machines} with {libFM}}.
\newblock \bibinfo{journal}{\emph{ACM Transactions on Intelligent Systems and
  Technology}} \bibinfo{volume}{3}, \bibinfo{number}{3} (\bibinfo{date}{May}
  \bibinfo{year}{2012}), \bibinfo{pages}{1--22}.
\newblock
\showISSN{2157-6904, 2157-6912}
\urldef\tempurl%
\url{https://doi.org/10.1145/2168752.2168771}
\showDOI{\tempurl}


\bibitem[Rendle(2013)]%
        {rendle2013scaling}
\bibfield{author}{\bibinfo{person}{Steffen Rendle}.}
  \bibinfo{year}{2013}\natexlab{}.
\newblock \showarticletitle{Scaling factorization machines to relational data}.
\newblock \bibinfo{journal}{\emph{Proceedings of the VLDB Endowment}}
  \bibinfo{volume}{6}, \bibinfo{number}{5} (\bibinfo{year}{2013}),
  \bibinfo{pages}{337--348}.
\newblock


\bibitem[Rendle et~al\mbox{.}(2011)]%
        {rendle2011fast}
\bibfield{author}{\bibinfo{person}{Steffen Rendle}, \bibinfo{person}{Zeno
  Gantner}, \bibinfo{person}{Christoph Freudenthaler}, {and}
  \bibinfo{person}{Lars Schmidt-Thieme}.} \bibinfo{year}{2011}\natexlab{}.
\newblock \showarticletitle{Fast context-aware recommendations with
  factorization machines}. In \bibinfo{booktitle}{\emph{Proceedings of the 34th
  international ACM SIGIR conference on Research and development in Information
  Retrieval}}. \bibinfo{pages}{635--644}.
\newblock


\bibitem[Roohi et~al\mbox{.}(2021)]%
        {roohi_predicting_2021}
\bibfield{author}{\bibinfo{person}{Shaghayegh Roohi},
  \bibinfo{person}{Christian Guckelsberger}, \bibinfo{person}{Asko Relas},
  \bibinfo{person}{Henri Heiskanen}, \bibinfo{person}{Jari Takatalo}, {and}
  \bibinfo{person}{Perttu Hämäläinen}.} \bibinfo{year}{2021}\natexlab{}.
\newblock \showarticletitle{Predicting {Game} {Engagement} and {Difficulty}
  {Using} {AI} {Players}}.
\newblock \bibinfo{journal}{\emph{Proceedings of the ACM on Human-Computer
  Interaction}} \bibinfo{volume}{5}, \bibinfo{number}{CHI PLAY}
  (\bibinfo{date}{Oct.} \bibinfo{year}{2021}), \bibinfo{pages}{1--17}.
\newblock
\showISSN{2573-0142}
\urldef\tempurl%
\url{https://doi.org/10.1145/3474658}
\showDOI{\tempurl}
\newblock
\shownote{arXiv: 2107.12061}.


\bibitem[Roohi et~al\mbox{.}(2020)]%
        {roohi2020predicting}
\bibfield{author}{\bibinfo{person}{Shaghayegh Roohi}, \bibinfo{person}{Asko
  Relas}, \bibinfo{person}{Jari Takatalo}, \bibinfo{person}{Henri Heiskanen},
  {and} \bibinfo{person}{Perttu H{\"a}m{\"a}l{\"a}inen}.}
  \bibinfo{year}{2020}\natexlab{}.
\newblock \showarticletitle{Predicting game difficulty and churn without
  players}. In \bibinfo{booktitle}{\emph{Proceedings of the Annual Symposium on
  Computer-Human Interaction in Play}}. \bibinfo{pages}{585--593}.
\newblock


\bibitem[Ryan et~al\mbox{.}(2006)]%
        {Ryan2006}
\bibfield{author}{\bibinfo{person}{Richard~M. Ryan}, \bibinfo{person}{C.~Scott
  Rigby}, {and} \bibinfo{person}{Andrew Przybylski}.}
  \bibinfo{year}{2006}\natexlab{}.
\newblock \showarticletitle{{The Motivational Pull of Video Games: A
  Self-Determination Theory Approach}}.
\newblock \bibinfo{journal}{\emph{Motivation and Emotion}}
  \bibinfo{volume}{30}, \bibinfo{number}{4} (\bibinfo{year}{2006}),
  \bibinfo{pages}{347--363}.
\newblock


\bibitem[Sarkar and Cooper(2017)]%
        {sarkar2017level}
\bibfield{author}{\bibinfo{person}{Anurag Sarkar} {and} \bibinfo{person}{Seth
  Cooper}.} \bibinfo{year}{2017}\natexlab{}.
\newblock \showarticletitle{Level difficulty and player skill prediction in
  human computation games}. In \bibinfo{booktitle}{\emph{Proceedings of the
  AAAI Conference on Artificial Intelligence and Interactive Digital
  Entertainment}}, Vol.~\bibinfo{volume}{13}. \bibinfo{pages}{228--233}.
\newblock


\bibitem[Silva et~al\mbox{.}(2015)]%
        {silva_dynamic_2015}
\bibfield{author}{\bibinfo{person}{Mirna~Paula Silva}, \bibinfo{person}{Victor
  do Nascimento~Silva}, {and} \bibinfo{person}{Luiz Chaimowicz}.}
  \bibinfo{year}{2015}\natexlab{}.
\newblock \showarticletitle{Dynamic {Difficulty} {Adjustment} through an
  {Adaptive} {AI}}. In \bibinfo{booktitle}{\emph{2015 14th {Brazilian}
  {Symposium} on {Computer} {Games} and {Digital} {Entertainment}
  ({SBGames})}}. \bibinfo{pages}{173--182}.
\newblock
\urldef\tempurl%
\url{https://doi.org/10.1109/SBGames.2015.16}
\showDOI{\tempurl}
\newblock
\shownote{ISSN: 2159-6662}.


\bibitem[Sweeney et~al\mbox{.}(2016)]%
        {sweeney_next-term_2016}
\bibfield{author}{\bibinfo{person}{Mack Sweeney}, \bibinfo{person}{Huzefa
  Rangwala}, \bibinfo{person}{Jaime Lester}, {and} \bibinfo{person}{Aditya
  Johri}.} \bibinfo{year}{2016}\natexlab{}.
\newblock \showarticletitle{Next-{Term} {Student} {Performance} {Prediction}:
  {A} {Recommender} {Systems} {Approach}}.
\newblock \bibinfo{journal}{\emph{arXiv:1604.01840 [cs]}}
  (\bibinfo{date}{Sept.} \bibinfo{year}{2016}).
\newblock
\urldef\tempurl%
\url{https://doi.org/10.5281/zenodo.3554603}
\showDOI{\tempurl}
\newblock
\shownote{arXiv: 1604.01840}.


\bibitem[Sweetser and Wyeth(2005)]%
        {sweetser2005gameflow}
\bibfield{author}{\bibinfo{person}{Penelope Sweetser} {and}
  \bibinfo{person}{Peta Wyeth}.} \bibinfo{year}{2005}\natexlab{}.
\newblock \showarticletitle{GameFlow: a model for evaluating player enjoyment
  in games}.
\newblock \bibinfo{journal}{\emph{Computers in Entertainment (CIE)}}
  \bibinfo{volume}{3}, \bibinfo{number}{3} (\bibinfo{year}{2005}),
  \bibinfo{pages}{3--3}.
\newblock


\bibitem[Tyack and Mekler(2020)]%
        {tyack2020self}
\bibfield{author}{\bibinfo{person}{April Tyack} {and} \bibinfo{person}{Elisa~D
  Mekler}.} \bibinfo{year}{2020}\natexlab{}.
\newblock \showarticletitle{{Self-Determination Theory in HCI Games Research:
  Current Uses and Open Questions}}. In \bibinfo{booktitle}{\emph{{Proc.
  Conference on Human Factors in Computing Systems (CHI)}}}. ACM,
  \bibinfo{pages}{1--22}.
\newblock


\bibitem[Van~Kreveld et~al\mbox{.}(2015)]%
        {kreveld2015automated}
\bibfield{author}{\bibinfo{person}{Marc Van~Kreveld}, \bibinfo{person}{Maarten
  L{\"o}ffler}, {and} \bibinfo{person}{Paul Mutser}.}
  \bibinfo{year}{2015}\natexlab{}.
\newblock \showarticletitle{Automated puzzle difficulty estimation}. In
  \bibinfo{booktitle}{\emph{2015 IEEE Conference on Computational Intelligence
  and Games (CIG)}}. IEEE, \bibinfo{pages}{415--422}.
\newblock


\bibitem[Weber and Notargiacomo(2020)]%
        {weber2020dynamic}
\bibfield{author}{\bibinfo{person}{Matheus Weber} {and}
  \bibinfo{person}{Pollyana Notargiacomo}.} \bibinfo{year}{2020}\natexlab{}.
\newblock \showarticletitle{Dynamic Difficulty Adjustment in Digital Games
  Using Genetic Algorithms}. In \bibinfo{booktitle}{\emph{2020 19th Brazilian
  Symposium on Computer Games and Digital Entertainment (SBGames)}}. IEEE,
  \bibinfo{pages}{62--70}.
\newblock


\bibitem[Wheat et~al\mbox{.}(2016)]%
        {wheat2016modeling}
\bibfield{author}{\bibinfo{person}{Daniel Wheat}, \bibinfo{person}{Martin
  Masek}, \bibinfo{person}{Chiou~Peng Lam}, {and} \bibinfo{person}{Philip
  Hingston}.} \bibinfo{year}{2016}\natexlab{}.
\newblock \showarticletitle{Modeling perceived difficulty in game levels}. In
  \bibinfo{booktitle}{\emph{Proceedings of the Australasian Computer Science
  Week Multiconference}}. \bibinfo{pages}{1--8}.
\newblock


\bibitem[Wilson et~al\mbox{.}(2019)]%
        {wilson2019eighty}
\bibfield{author}{\bibinfo{person}{Robert~C Wilson}, \bibinfo{person}{Amitai
  Shenhav}, \bibinfo{person}{Mark Straccia}, {and} \bibinfo{person}{Jonathan~D
  Cohen}.} \bibinfo{year}{2019}\natexlab{}.
\newblock \showarticletitle{The eighty five percent rule for optimal learning}.
\newblock \bibinfo{journal}{\emph{Nature communications}} \bibinfo{volume}{10},
  \bibinfo{number}{1} (\bibinfo{year}{2019}), \bibinfo{pages}{1--9}.
\newblock


\bibitem[Xue et~al\mbox{.}(2017)]%
        {xue_dynamic_2017}
\bibfield{author}{\bibinfo{person}{Su Xue}, \bibinfo{person}{Meng Wu},
  \bibinfo{person}{John Kolen}, \bibinfo{person}{Navid Aghdaie}, {and}
  \bibinfo{person}{Kazi~A. Zaman}.} \bibinfo{year}{2017}\natexlab{}.
\newblock \showarticletitle{Dynamic {Difficulty} {Adjustment} for {Maximized}
  {Engagement} in {Digital} {Games}}. In \bibinfo{booktitle}{\emph{Proceedings
  of the 26th {International} {Conference} on {World} {Wide} {Web} {Companion}
  - {WWW} '17 {Companion}}}. \bibinfo{publisher}{ACM Press},
  \bibinfo{address}{Perth, Australia}, \bibinfo{pages}{465--471}.
\newblock
\showISBNx{978-1-4503-4914-7}
\urldef\tempurl%
\url{https://doi.org/10.1145/3041021.3054170}
\showDOI{\tempurl}


\bibitem[Zook et~al\mbox{.}(2012)]%
        {zook_skill-based_2012}
\bibfield{author}{\bibinfo{person}{Alexander Zook}, \bibinfo{person}{Stephen
  Lee-Urban}, \bibinfo{person}{Michael~R. Drinkwater}, {and}
  \bibinfo{person}{Mark~O. Riedl}.} \bibinfo{year}{2012}\natexlab{}.
\newblock \showarticletitle{Skill-based {Mission} {Generation}: {A}
  {Data}-driven {Temporal} {Player} {Modeling} {Approach}}. In
  \bibinfo{booktitle}{\emph{Proceedings of the {The} third workshop on
  {Procedural} {Content} {Generation} in {Games} - {PCG}'12}}.
  \bibinfo{publisher}{ACM Press}, \bibinfo{address}{Raleigh, NC, USA},
  \bibinfo{pages}{1--8}.
\newblock
\showISBNx{978-1-4503-1447-3}
\urldef\tempurl%
\url{https://doi.org/10.1145/2538528.2538534}
\showDOI{\tempurl}


\end{thebibliography}

\end{document}